# Galactic Evolution of Silicon Isotopes:
# Application to Presolar SiC Grains From Meteorites

F. X. Timmes[1,2] and D. D. Clayton[1]

fxt@burn.uchicago.edu and clayton@gamma.phys.clemson.edu

[1] Department of Physics and Astronomy
Clemson University
Clemson, SC   29634

[2] Enrico Fermi Institute
Laboratory for Astrophysics and Space Research
University of Chicago
Chicago, IL   60637

# ABSTRACT

We calculate and discuss the chemical evolution of the isotopic silicon abundances in the interstellar medium at distances and times appropriate to the birth of the solar system. This has several objectives, some of which are related to anomalous silicon isotope ratios within presolar grains extracted from meteorites; namely: (1) What is the relative importance for silicon isotopic compositions in the bulk ISM of Type II supernovae, Type Ia supernovae, and AGB stars? (2) Are $^{29}$Si and $^{30}$Si primary or secondary nucleosynthesis products? (3) In what isotopic direction in a three-isotope plot do core-collapse supernovae of different mass move the silicon isotopic composition? (4) Why do present calculations not reproduce the solar ratios for silicon isotopes, and what does that impose upon studies of anomalous Si isotopes in meteoritic silicon carbide grains? (5) Are chemical-evolution features recorded in the anomalous SiC grains? Our answers are formulated on the basis of the Woosley & Weaver (1995) supernova yield survey. Renormalization with the calculated interstellar medium silicon isotopic composition and solar composition is as an important and recurring concept of this paper. Possible interpretations of the silicon isotope anomalies measured in single SiC grains extracted from carbonaceous meteorites are then presented. The calculations suggest that the temporal evolution of the isotopic silicon abundances in the interstellar medium may be recorded in these grains.

Subject headings: ISM: Abundances – Nuclear Reactions, Nucleosynthesis, Abundances – Supernovae: General – Dust, Extinction

# 1. INTRODUCTION

We have two motivations for studying the silicon isotopic abundance histories in the interstellar medium at distances and times appropriate to the birth of the solar system. Primarily we seek connections between evolution of the calculated silicon isotopic ratios and the anomalous silicon isotopic ratios found in presolar silicon carbide grains extracted from meteorites. Secondarily we assess contributions to the silicon isotopes from the various sources that produce them. Individual grains which condensed from stellar outflows and migrated into the solar nebula, found today in carbonaceous meteorites, has opened unique views on stellar nucleosynthesis, star-formation processes, local mixing processes in the interstellar medium (henceforth ISM), and chemical evolution (Clayton 1982). The most clearcut cases involve grains possessing such large isotopic anomalies that they surely formed within ejecta from specific stars prior to mixing with the ISM. The morphology and composition of these presolar grains have been reviewed by Anders & Zinner (1993) and by Ott (1993).

The first stardust grains to be isolated were of carbonaceous composition, specifically diamonds, silicon carbide (henceforth SiC) and graphite (Anders & Zinner 1993). Acid-resistant residues of carbonaceous meteorites had already been shown in the early 1960's to be isotopically anomalous in their xenon content known as Xe-HL (Reynolds & Turner 1964). Other xenon enrichments were recognized to be displaying a crisp s-process signature (Clayton & Ward 1978; Srinivasan & Anders 1978). Subsequent isolation, purification, and characterization of that acid-resistant residue allowed its identification as SiC (Bernatowicz et al. 1987; Tang & Anders 1988). Isotopic studies of not only the noble gases but also carbon, silicon, and other trace elements with secondary ion mass spectrometers led to the clear identification of huge isotopic anomalies in SiC (Lewis et al. 1991; Anders & Zinner 1993 and references therein). The anomalous isotopes and the almost pure s-process xenon mark these presolar SiC grains as having formed from ejecta rich in the nucleosynthesis products of a single star. The name STARDUST has been suggested for high quality single grains grown in stellar winds, to distinguishing them from other anomalous samples, and a related name, SUNOCON, labels supernova condensates grown in the ejecta before it is mixed with the ISM (Clayton 1978).

Stellar nucleosynthesis modeling has been concerned chiefly with reproducing the bulk composition of the solar system, an important concern in its own right, but individual grains which are isotopically anomalous yield information about very specific stellar origin sites. For SiC grains, an origin site where silicon and carbon can condense without being significantly oxidized seems necessary. The consensus picture which has taken shape is that SiC grains condense in the outflows from intermediate and low mass stars when they enter the carbon star phase. Carbon stars being defined as asymptotic giant branch (henceforth AGB) stars in which the atmospheric carbon to oxygen ratio is greater than unity, and molecular band heads of $C_2$ are prominent in the spectra. This paradigm is built on the careful isolation and characterization of presolar SiC grains by groups at Bern, Caltech, Chicago, and St. Louis (Zinner, Tang, & Anders 1987, 1989; Tang et al. 1989; Stone et al. 1991; Lewis et al. 1991; Virag et al. 1992; Alexander 1993; Hoppe et al. 1994ab, 1996; Nittler et al. 1995ab, 1996). We have progressed from purely theoretical predictions (Clayton 1978) to having found such grains in meteorites, and thus can study individual pieces of individual stars in the laboratory.

Additional and independent evidence in favor of a carbon star origin site is that quite a few s-process elements have been observed to be enriched in the atmospheres of carbon stars (Sneden & Parthasarathy 1983; Luck & Bond 1985; Sneden & Pilachowski 1985, Gilroy et al. 1988; Sneden et al. 1988; Gratton & Sneden 1994; Cowan et al. 1995). Mixing processes in the ISM would have destroyed and severely diluted the s-process Xe found in grains had the xenon not been trapped by the grains during outflows from carbon stars. Barium and neodymium s-process isotopes have also been found in SiC grains (Ott & Begemann 1990, Zinner, Amari, & Lewis 1991; Prombo et al 1993; Richter et al. 1992, 1993).

An issue addressed in this paper is whether silicon isotopic anomalies in presolar SiC grains are to be interpreted exclusively in terms of the nucleosynthesis from individual stars, or whether some effects may be due the chemical evolution of the matter from which the individual stars form (Clayton 1988; Clayton, Scowen & Liffman 1989; Alexander 1993). Not all anomalous SiC grains are clearly attributable to carbon star condensate. The class of SiC grains known as X grains bear large $^{29}$Si and $^{30}$Si deficits, most are rich in $^{12}$C with $^{12}$C/$^{13}$C ratios approaching ten times the solar ratio, and contain large excesses of $^{49}$Ti and $^{44}$Ca. (Amari et al. 1992; Nittler et al. 1995ab, 1996). Clayton (1975, 1978, 1981) predicted excess $^{49}$Ti and $^{44}$Ca within SUNOCONs, owing to condensation of radioactive $^{49}$V and $^{44}$Ti within expanding supernova ejecta and the $^{12}$C-rich helium burning shell. The confirmed existence of such isotopic effects (Nittler et al. 1995ab, 1996) lends strong support for assuming these grains are SUNOCONs. All this does not, however, eliminate problems in interpreting the silicon isotope ratios measured in X grains. The question is whether the bulk silicon ejecta which condenses into X grains is sufficiently enriched in $^{28}$Si (roughly twofold with respect to $^{29,30}$Si), and relatively richer in $^{29}$Si than in $^{30}$Si.

In the literature it is conventional to express the silicon (and other element) isotope ratios in parts per thousand deviation from the solar silicon isotope ratio:

$$\delta_\odot \left(\frac{^{29}\text{Si}}{^{28}\text{Si}}\right) \equiv 1000 \left[\left(\frac{^{29}\text{Si}}{^{28}\text{Si}}\right) / \left(\frac{^{29}\text{Si}}{^{28}\text{Si}}\right)_\odot - 1\right] \qquad \delta_\odot \left(\frac{^{30}\text{Si}}{^{28}\text{Si}}\right) \equiv 1000 \left[\left(\frac{^{30}\text{Si}}{^{28}\text{Si}}\right) / \left(\frac{^{30}\text{Si}}{^{28}\text{Si}}\right)_\odot - 1\right]. \qquad (1)$$

For ease of notation (and reading) these will be denoted as $\delta_\odot^{29}$Si and $\delta_\odot^{30}$Si, respectively. It is traditional to use these definitions in a "three-isotope plot"; $\delta_\odot^{29}$Si versus $\delta_\odot^{30}$Si in a Cartesian plane. The silicon isotopic composition of any SiC grain is represented by a single point in a three-isotope plot. Two silicon isotope compositions form two points, and any linear combination of these two compositions lies along the line connecting those two points.

It will prove useful to distinguish between an ISM normalization as well as a solar normalization, since computed ISM silicon isotopic abundances may not pass precisely through solar silicon:

$$\delta_{\text{ISM}} \left(\frac{^{29}\text{Si}}{^{28}\text{Si}}\right) \equiv 1000 \left[\left(\frac{^{29}\text{Si}}{^{28}\text{Si}}\right) / \left(\frac{^{29}\text{Si}}{^{28}\text{Si}}\right)_{\text{ISM}} - 1\right] \qquad \delta_{\text{ISM}} \left(\frac{^{30}\text{Si}}{^{28}\text{Si}}\right) \equiv 1000 \left[\left(\frac{^{30}\text{Si}}{^{28}\text{Si}}\right) / \left(\frac{^{30}\text{Si}}{^{28}\text{Si}}\right)_{\text{ISM}} - 1\right], \qquad (2)$$

and will be compactly denoted as $\delta_{\text{ISM}}^{29}$Si and $\delta_{\text{ISM}}^{30}$Si, respectively.

Measurement of the silicon isotopes in most SiC grains demonstrate that both $\delta^{29}_\odot$Si and $\delta^{30}_\odot$Si are larger than zero (Zinner, Tang, & Anders 1987, 1989; Stone et al. 1991; Virag et al. 1992; Alexander 1993; Hoppe et al. 1994ab). That is, mainstream SiC grains are enriched in $^{29}$Si and $^{30}$Si relative to solar. More surprising is that $\delta^{29}_\odot$Si correlates strongly , grain for grain, with $\delta^{30}_\odot$Si along a best-fit line of slope 1.34 (Hoppe et al. 1994a). There is no corresponding correlation in the carbon isotopes, which are highly variable (Zinner et al. 1987; 1989; Stone et al. 1991; Virag et al. 1992; Alexander 1993). This requires the stellar origin sites to preferentially affect carbon isotopic ratios rather than silicon isotopic ratios. Grain condensation in the winds of carbon stars becomes an even more attractive hypothesis since the cumulative effects of dredgeup, mass loss, and hot-bottom burning can produce the widely varying carbon isotopic compositions that is observed in solar vicinity giants (Lambert et al. 1986), while scarcely affecting the silicon isotopic composition.

Only neutron capture reactions are expected to modify the silicon isotopic composition in AGB stars. One expects s-processing in the helium shell, interspersed with dredgeups, to show monotonically evolving silicon isotopic ratios. A carbon star origin site for presolar SiC grains would almost be regarded as settled were it not for the fact that s-processing of the silicon isotopes produces a $\delta^{29}_\odot$Si $-$ $\delta^{30}_\odot$Si correlation with a slope of 0.46 instead of the measured slope of 1.34. Neutron fluxes always produce more excess $^{30}$Si than $^{29}$Si because of their relative neutron capture cross sections, and because of the large $^{33}$S(n,$\alpha$)$^{30}$Si cross section (Bao & Käppeler 1987; Brown & Clayton 1992). This forces one to reach deeper for a satisfactory explanation, perhaps even casting some doubt on the hypothesis of a carbon star origin. The puzzle drove Brown & Clayton (1992) to propose that only the most massive AGB stars, whose helium shell thermal flashes are hot enough for $\alpha$ reactions on magnesium isotopes to occur, could condense presolar SiC grains. They showed that in this case a slope of 1.34 for the evolution of the surface composition was at least a technical possibility, if an improbable one, in individual AGB stars. The correlations of $\delta^{46}_\odot$Ti with $\delta^{30}_\odot$Si, however, fairly convincingly rule out this possibility (Hoppe et al. 1994ab).

The rate of occurrence of AGB stars is quite high relative to the number of massive stars, although the number visible at any given time is modest owing to their rapid evolution through the AGB phase. If micron sized SiC grains live, on average, several 100 Myr in the ISM before being incorporated into a molecular cloud, it is not hard to see that many AGB stars could have contributed to the presolar SiC grain population. A simple order-of-magnitude estimate (Alexander 1993) for the number $N_{\rm AGB}$ of AGB stars that pass through a molecular cloud is given by the product of the mean number of AGB stars that form throughout the Galaxy during the lifetime of the molecular cloud and the volume fraction of the Galaxy occupied by the molecular cloud:

$$N_{\rm AGB} = R_{\rm PN}\ T\ M\ V_{\rm G}/M_{\rm G}\ , \qquad (3)$$

where $R_{\rm PN} \approx 3$ yr$^{-1}$ is the average formation rate of white dwarfs / planetary nebula in the Galaxy, $T \approx 10^8$ is the mean lifetime of an individual molecular cloud, $M$ is the mass of an individual molecular cloud, $V_{\rm G} \approx 2.5 \times 10^{-4}$ is the fractional volume occupied by the sum of all molecular clouds in the Galaxy, and $M_G \approx 10^9$ M$_\odot$ is the total mass of all molecular clouds in the Galaxy (Alexander 1993). For a molecular cloud mass of $10^6$ M$_\odot$, $N_{\text{AGB}} \approx 75$. An astrophysically interesting variation of this estimate is that a $10^6$ M$_\odot$ ISM mass probably spends half its time at a number density of $\simeq 10^3$ cm$^{-3}$ and half its time at the ambient $\simeq 1$ cm$^{-3}$. The average cloud volume is then 500 times larger than the volume assumed above, which propagates into 500 times more AGB stars seeding a cloud with SiC grains. Should SiC grains survive longer than the $10^8$ yr lifetime of a cloud (say $10^9$ yr), another factor of ten is gained, and the number of AGB stars seeding a cloud with SiC grains is 5000 times greater than the canonical estimate given above. Either way, the most probable value of $N_{\text{AGB}}$ suggests many AGB stars could seed a large molecular cloud with SiC grains. Turbulence within the cloud may or may not be needed to spatially distribute the grains, depending on the value of $N_{\text{AGB}}$.

Under a "many-AGB-star" hypothesis (Alexander 1993; Gallino et al. 1994), variations should exist in the initial compositions of stars owing, for example, to continuing star formation or ISM mixing processes. Since abundances of the primary ($^{28}$Si) and secondary ($^{29,30}$Si) isotopes grow at different rates in mean chemical evolution models, older stars are, on average, more deficient in the secondary isotopes. A collection of SiC grains could distribute their silicon isotopic compositions along a line in a three-isotope plot if the AGB initial silicon isotopic compositions lay along a line. Nor are AGB stars the only potential sources for SiC. Wolf-Rayet carbon winds and post-supernova helium shells of massive stars provide other potential sources for SiC grains. As an example, two WC stars could have distinct initial compositions owing to differential enrichment by prior supernovae that triggered their formation.

Evaluation of any of these options requires an understanding of the mean chemical evolution of the silicon isotopes. After first examining the nucleosynthesis of silicon isotopes from massive stars, Type Ia supernova, and AGB stars in §2, (noting an exceptional situation in §2.4), we delineate the mean chemical evolution in the solar vicinity, mean injection rates into the ISM in the solar neighborhood, and signatures due to incomplete mixing in §3. The mainstream SiC rains occupies much §4, with a possible interpretation of them given in §4.6, and a potential solution to the silicon isotope ratios measured in X-type SiC grains given in §4.7. After surveying the available evidence and inferences, consideration is given to problems that may still remain in our current understanding of the anomalous silicon isotope ratios in presolar SiC grains from meteorites.

## 2. NUCLEOSYNTHESIS OF THE SILICON ISOTOPES

Type II supernovae are the principle origin site of the vast majority of the chemical elements, including the silicon isotopes. Typically, the matter ejected contains about 10 times as many atoms of a given heavy element than did the initial matter of the massive star. Type Ia supernovae can affect the evolution of the silicon isotopes by several percent. AGB stars may also inject small, but interesting, amounts of silicon into the ISM under certain conditions. Hydrostatic oxygen burning, explosive carbon, oxygen or neon burning, and slow neutron captures are the general processes that change silicon isotopic composition in stars. In the remainder of this section the nucleosynthesis of silicon from these various sources and processes are discussed.

## 2.1 TYPE II AND Ib SUPERNOVA

Figure 1a shows the $^{28}$Si yields from the supernova models of Woosley & Weaver (1995). The points labeled with the symbol "u" represent stars with an initial metallicity of $10^{-4}$ $Z_\odot$, "t" for $10^{-2}$ $Z_\odot$, "p" for 0.1 $Z_\odot$, and "s" for 1.0 $Z_\odot$. These $^{28}$Si yields are not monotonic with respect to stellar mass. Variations are caused by differences in the density structure of the pre-supernova stars, the sensitivity of the pre-supernova models to the interaction of the various convective zones during oxygen and silicon burning, the uncertainty in modeling the explosion mechanism, and the mass of freshly synthesized silicon which may fall-back onto the compact remnant. However, these small variations overlie a fundamental property; namely, that production of $^{28}$Si proceeds just as easily from a star composed primarily of hydrogen and helium (points u) as it does in massive stars with a much larger initial metallicity (points s). Fig. 1a shows that production of $^{28}$Si is "primary" – a term reserved for isotopes whose production is generally independent of the initial metallicity of the star.

The same statements are not true for the heavier stable isotopes of silicon. Figures 1b and 1c show the $^{29}$Si and $^{30}$Si yields, respectively, on a logarithmic ordinate. The labels (u, t, p, s) have the same meaning as above, and the $^{29,30}$Si yields vary with stellar mass for most of the same reasons as does $^{28}$Si. However, they are not as influenced by fall-back since $^{29,30}$Si are chiefly synthesized farther out from the core than $^{28}$Si. This explains why the yields shown in Figs. 1b and 1c don't decline at larger stellar masses as they do in Fig. 1a. The important point is that $^{29,30}$Si yields strongly depend on the initial metallicity of the massive star, i.e., they are "secondary". The ejected masses for these neutron-rich isotopes increase with the initial metallicity of the massive stars (s > p > t > u).

It is instructive to recall how the extra neutrons that allow the production of $^{29}$Si and $^{30}$Si in post helium burning processes are released. The neutron excess $\eta$ is defined as

$$\eta = \sum (N_i - Z_i) Y_i \ , \qquad (4)$$

where $N_i$ is the number of neutrons in species $i$, $Z_i$ is the number of protons, and $Y_i$ is the normalized ($\sum Y_i$=1) molar abundance. A pure proton composition has $\eta$ = -1, matter with an equal number of protons and neutrons has $\eta$=0, while a pure neutron composition has $\eta$ = 1.

Hydrogen burning on the main-sequence transforms carbon and oxygen into $^{14}$N. Two successive $\alpha$-particle captures on $^{14}$N during core helium burning produces the classic $^{22}$Ne neutron source: $^{14}$N($\alpha,\gamma$)$^{18}$F($e^+\nu$)$^{18}$O($\alpha,\gamma$)$^{22}$Ne ($\alpha$,n)$^{25}$Mg. The two isotopes $^{29,30}$Si are then synthesized mainly through $^{23}$Na($\alpha$,p)$^{26}$Mg($\alpha$,n)$^{29}$Si, $^{28}$Si(n,$\gamma$)$^{29}$Si(n,$\gamma$)$^{30}$Si and $^{24}$Mg($\alpha$,p)$^{27}$Al($\alpha$,p)$^{30}$Si (Pardo, Couch & Arnett 1974; Thielemann & Arnett 1985; Woosley & Weaver 1982; 1995). These reactions occur partly during the hydrostatic helium and carbon burning phases, but mostly during shell oxygen burning. Before the production of $^{22}$Ne the neutron excess is essentially zero in the zones where hydrogen as been burned, while after the $^{22}$Ne($\alpha$,n)$^{25}$Mg reaction $\eta \simeq 0.0019$ $Z/Z_\odot$ (Woosley & Weaver 1982). The exact distribution of $^{22}$Ne within the massive star is important, but is overshadowed by the fact that $^{22}$Ne is produced is proportion to the initial CNO content of the star. Hence, as the initial metallicity of the star increases, yields of $^{29,30}$Si increase.

The situation is actually more complicated than a simple initial CNO dependence. Weak interactions during post helium burning phases can substantially alter the neutron excess (Thielemann & Arnett 1985). This decreases the strict dependence on the initial metallicity. For example, massive stars having an initial metallicity $Z \leq 0.1$ $Z_\odot$ build up a small neutron excess ($\simeq 3.7 \times 10^{-4}$) which is independent of the initial metallicity (Woosley & Weaver 1982). This effect can be discerned in Figs. 1b and 1c in two ways. First, by the close similarity of the $^{29,30}$Si yields from the low-metallicity stars (points u and t). Second, the yields are not strict multiples of each other; solar metallicity yields are not simply 10 times the 0.1 $Z_\odot$ yields. Shell oxygen burning, which is the location of the freshly minted silicon that can escape from the star, occurs at lower density than core oxygen burning. As such, weak decays interactions during shell burning are less important than during core burning. Despite all these complications about the amplitude and distribution of $\eta$, it remains true that the heavy silicon isotopes are a secondary nucleosynthetic product.

Location of the silicon isotopic compositions in the Woosley & Weaver (1995) models, and all the other isotopes, are conveniently expressed in Meyer, Weaver & Woosley (1995). While zone compositions of Type II supernovae are relevant for SUNOCONs (Clayton 1978), in the present paper only the bulk composition of supernova ejecta is considered.

Another effort to model nucleosynthesis in massive stars in detail commensurate to the Woosley & Weaver (1995) survey is Thielemann, Nomoto & Hashimoto (1996). They find silicon yields that are sometimes similar, sometimes not, to the Woosley & Weaver (1995) values. A discussion of the reasons for the differences between the two studies is given by Woosley & Weaver (1995). For our purposes here, it is sufficient to note that when the Thielemann et al. nuclear reaction rates are used in the Woosley & Weaver stellar models, the differences in the silicon yields are less than 0.1% (Hoffman et al. 1996). This level of agreement ensues chiefly because the two groups use the same experimentally determined $^{28,29,30}$Si(p,$\gamma$) and $^{28,30}$Si($\alpha$,n) reaction rates. The rest of the rates that affect silicon production originate from theoretical Hauser-Feshbach calculations, and differences there do not appear to significantly affect the yields. Thus, the main reasons for the differences in the silicon yields between the two groups are tied to the different adiabatic paths followed in the explosion and the progenitor structure (Hoffman et al. 1996).

Convective oxygen shell-burning prior to core collapse in a 20 $M_\odot$ star was examined in two dimensions by Bazan & Arnett (1994). They find plume structures dominate the velocity field, and that significant mixing beyond the boundaries defined by mixing-length theory (i.e., "convective overshoot") brings fresh fuel (carbon) into the convective region. This causes local hot spots of nuclear burning. This general picture is dramatically different from the one-dimensional situation. While no yields from 2D calculations are presently available, it is likely that local burning and chemical inhomogeneities will change the silicon isotope yields from a single supernova. However, integration over an initial mass function smoothes out stochastic yields from stars of different mass or even different yields from the same progenitor mass (e.g., Arnett 1995). Thus, the general features of mean chemical evolution as determined from one dimensional stellar models, may remain intact. Factors of two variation in the yields from individual supernovae, however, can be quite significant for meteoritic grains which may originate from inhomogeneous enrichments of stars.

Silicon isotope ejecta from the solar metallicity Type II supernovae models are shown in the three-isotope diagram in Figure 2, and listed in the middle two columns of Table 1. These are raw ratios of the total isotopic mass ejected, unnormalized to any reference composition. Type II supernova yields depend on the initial metallicity, but not on the initial silicon content of the progenitor. That is, the silicon isotopic ratios ejected from a given supernova is independent of the initial silicon isotopic ratios. Fig. 2 can then be taken to represent massive star ejecta for all solar CNO initial compositions. Two special points are shown in Fig. 2. The first is the solar silicon composition. Note that it is not reproduced by any solar metallicity supernova. The second special point (marked with the large "+" symbol) is the silicon isotopic ratios in the ISM when the Sun was born, as calculated from the mean chemical evolution model to be discussed in §3. Note it is not equal to solar.

The $^{29}$Si=$^{30}$Si line drawn in Fig. 2 shows that all these supernovae models eject roughly equal masses of $^{29}$Si and $^{30}$Si. This is the result of a complex interplay between thermal conditions, convection and nuclear reactions rates (see §2.4). It should not be surprising then when we show in §3 that mean chemical evolution models, which are dominated by the ejecta of core collapse events, produce m=1 slope lines in a $\delta$-value three-isotope plot, when the evolutions are normalized with respect to the calculated mean ISM composition at solar birth. The slope would not be unity if a solar composition was used for as the reference point. This crucial point is analyzed in detail in §4.6. However, a slope one line when absolute silicon isotopic ratios are plotted should not be confused with a slope one line in a three-isotope plot since they are very different quantities with different properties. It is to this bewildering array of silicon compositions that order is sought.

## 2.2 TYPE Ia SUPERNOVAE

The standard model for Type Ia supernova consists of a carbon-oxygen white dwarf that accretes mass from a binary companion at the proper rate for a sufficient time such that it grows to nearly the Chandrasekhar mass (1.39 $M_\odot$), at which point it ignites carbon near the center. The successes and failures of this model in reproducing observed Type Ia light curves and spectral properties has been discussed extensively. Production of silicon follows essentially the same pathways as for core collapse supernovae, but there may be large differences due to electron capture occuring at higher densities for longer periods of time. For example, various models eject different silicon to iron ratios due to various assumptions of how much material experiences how much electron capture for how long (Thielemann, Nomoto & Yokoi 1986; Woosley & Weaver 1993; Khokhlov 1993; and Arnett & Livne 1994). These assumptions, in turn, govern the global evolutionary properties of the Chandrasekhar mass white dwarf (e.g., outright explosion, or expansion first, collapse, and then explosion).

The W7 model of Thielemann et al. (1986) is adopted as representative of Type Ia supernovae nucleosynthesis. Most of the $^{28}$Si ejected by W7 comes from incomplete silicon burning between $0.75 \leq M/M_\odot \leq 1.0$, and explosive oxygen and neon burning in the outer layers. Explosive carbon burning in the outer layers mainly produces $^{20}$Ne, but is also produces most of the $^{29}$Si and $^{30}$Si. W7 has an initial composition of equal $^{12}$C and $^{16}$O mass fractions and a supersolar $^{22}$Ne mass fraction of 0.025. W7 ejects 0.15 $M_\odot$ of $^{28}$Si, $3.0\times10^{-4}$ $M_\odot$ of $^{29}$Si, and $3.4\times10^{-3}$ $M_\odot$ of $^{30}$Si. A potential concern for bulk silicon isotope evolution is sensitivity of the Type Ia yields to the initial composition. Early on in the Galaxy's evolution when very low-metallicity massive stars are becoming Type II supernovae, chemical evolution models which uniformly apply W7 will slightly overestimate the $^{29,30}$Si contributions from Type Ia events. Uniform application of W7 does not introduce a large error later in the Galaxy's evolution (e.g., birth of the Sun) since by then Type II supernovae have, and continue, to dictate both the absolute abundance levels and the injection rates of the silicon isotopes (see §3).

There are several poorly understood aspects of the standard Type Ia supernova model. How is the nova instability suppressed if the white dwarf slowly accretes hydrogen-rich material? Why is the central region ignited, rather than off center or near the edge if two carbon-oxygen white dwarfs are merging? What physics controls the flame propagation such that the overproduction of rare neutron-rich isotopes ($^{54,58}$Fe, $^{545}$Cr, $^{58}$Ni) does not occur? Where are the white dwarf progenitors from an observational standpoint? Sufficient uncertainty exists to warrant investigation into alternative models (Woosley & Weaver 1994).

Stellar evolution studies suggest that common 0.6 – 0.9 $M_\odot$ CO white dwarfs that merge with a helium main-sequence star, accreting helium at a rate of several times $10^{-8}$ $M_\odot$ yr$^{-1}$, may be an attractive Type Ia supernovae model (e.g., Tutukov, Yungelson, & Iben 1992). When 0.15 – 0.20 $M_\odot$ of helium has been accreted, a detonation is initiated at the base of the accreted layer. This helium detonation compresses the CO material and triggers a detonation of the core (Livne & Glasner 1991; Woosley & Weaver 1994).

Behavior of the silicon isotopes in the Chandrasekhar mass W7 model and two representative sub-Chandrasekhar Type Ia models are shown in Figure 3. The upper portion of the figure gives the total ejected silicon masses, while the lower portion gives the ejected mass fractions divided by the appropriate solar mass fraction. Fig. 3 is further divided into three vertical sections, one for W7 (Thielemann et al. 1986), one for a 0.6 $M_\odot$ sub-Chandrasekhar model, and one for a 0.9 $M_\odot$ sub-Chandrasekhar model (Woosley & Weaver 1994). These latter two models are representative of the range encompassed by sub-Chandrasekhar mass Type Ia models. Model 1 accretes 0.2 $M_\odot$ of helium and ejects 0.27 $M_\odot$ of $^{56}$Ni, 0.14 $M_\odot$ of $^{28}$Si, $5.0\times10^{-5}$ $M_\odot$ of $^{29}$Si and $7.8\times10^{-5}$ $M_\odot$ of $^{30}$Si. Model 8 also accretes 0.2 $M_\odot$ of helium but ejects 0.79 $M_\odot$ of $^{56}$Ni, $7.8\times10^{-2}$ $M_\odot$ of $^{28}$Si, $5.5\times10^{-5}$ $M_\odot$ of $^{29}$Si, and $7.2\times10^{-5}$ $M_\odot$, of $^{30}$Si.

Note that all the Type Ia models in Fig. 3 underproduce the neutron-rich silicon isotopes in comparison to $^{28}$Si, even for W7 with it's large initial $^{22}$Ne mass fraction. It is this feature that makes contributions to $^{29,30}$Si from Type Ia events unimportant for bulk Galactic material (see §3). As far as the evolution of the silicon isotopes is concerned, the exact nature of Type Ia progenitors matters little.

## 2.3 INTERMEDIATE AND LOW MASS STARS

In principle, several nuclear processes can change the silicon isotopic ratios in intermediate and low mass stars. In mild hydrogen burning, where the temperature ranges from $8 - 10\times10^7$ K, proton captures on $^{27}$Al create $^{28}$Si. This form of burning can occur in some hot bottom burning models at the base of the convective envelope for stars more massive than $\simeq$ 5 M$_\odot$. In fast hydrogen burning, where the temperature exceeds $1\times10^8$ K, proton captures destroy more $^{29}$Si present than either $^{28}$Si or $^{30}$Si. This process can occur at the base of the convective envelope for stars lighter than $\simeq$ 7 M$_\odot$. The s-process can occur in the helium burning shell of thermally pulsing AGB stars, provided the $^{13}$C or $^{22}$Ne neutron source is present, and produces comparable masses of $^{29,30}$Si. During core helium burning, where the temperature exceeds $4\times10^8$ K for a sufficiently long time, $\alpha$-captures on $^{12}$C can produce $^{28}$Si. Production of $^{28}$Si by this process in thermally pulsing AGB stars depends sensitively on thermodynamic conditions. "Magnesium burning", where $\alpha$ particles capture on $^{25,26}$Mg to produce $^{29,30}$Si respectively, can occur if the He-shell peak temperature reaches $450\times10^6$ K. The magnesium isotopes may be present in the initial composition of the intermediate/low mass star, or be made in-situ by the s-process. Details of these processes are discussed in Brown & Clayton (1992).

In published hot bottom burning models the temperatures are $\simeq 50\times10^6$ K; too small to have significant proton capture reactions on silicon in the envelope. Boothroyd, Sackmann & Wasserburg (1995) reported peak temperatures at the base of the envelope of $70\times10^6$ K in their 5 M$_\odot$ star, barely reaching $100\times10^6$ K in the 7 M$_\odot$ star. These stars do not spend a long enough time in the AGB phase or experience as many thermal pulses, so that the shell burning temperatures are limited to $\approx 3\times10^8$ K. Thus, it appears likely that only the s-process can make substantial changes to the silicon isotopic ratios in AGB stars.

Evolution of the silicon isotopes due to s-processing in AGB stars can be estimated in a simple way. Consider two propositions: (a) the total mass ejected over the star's lifetime, not just during the carbon star phase, is composed of 90% initial envelope material plus 10% of material dredged up from the helium shell, and (b) the $^{29}$Si and $^{30}$Si mass fractions in the helium shell are enriched by 40% and 87%, respectively, when normalized to solar. Both these propositions have been substantiated by several investigations (Brown & Clayton 1992; Gallino et al. 1994). Ignoring small changes in $^{28}$Si so that $^{28}$Si$_\mathrm{out}$ = $^{28}$Si$_\mathrm{in}$, the superposition of propositions (a) and (b) for solar metallicity stars gives the normalized excesses produced by the s-process:

$$^{29}\mathrm{Si}_\mathrm{out} = 1.040 \ ^{29}\mathrm{Si}_\mathrm{in} \qquad ^{30}\mathrm{Si}_\mathrm{out} = 1.087 \ ^{30}\mathrm{Si}_\mathrm{in} \ . \tag{5}$$

It is important the production factor ratio (0.04/0.087) always remain at the s-process value 0.46.

The broad peak in the G-dwarf distribution of solar vicinity stars suggests that AGB stars born with metallicities around 0.4 Z$_\odot$ could have been a common contributor to the presolar ISM. This depends, of course, on the initial mass function and the evolutionary timescales to ascend to the AGB phase as a function of the initial stellar mass. The normalized excesses (but not the production factor ratio) in eq. (6) will be larger for AGB stars with smaller initial metallicities. Why will they be larger? The neutron fluxes in the interpulse mixing pocket should be adequate to drive the silicon isotopes into flow equilibrium. Thus, the mass fractions of $^{29,30}$Si ejected is a certain percentage of the initial $^{28}$Si mass fraction, independent of the small initial $^{29,30}$Si mass fractions. Hence, the normalized excesses are (slightly) larger than indicated by eq. (5). In addition, all stars relevant to the presolar ISM will begin their lives with roughly a solar ratio of the $\alpha$-chain elements, $^{28}$Si/$^{32}$S for example. Significant $^{30}$Si production then occurs through $^{32}$S(n,$\gamma$)$^{33}$S(n,$\alpha$)$^{30}$Si, and so the small initial $^{30}$Si mass fraction is quickly forgotten.

The arguments above advocate $^{29,30}$Si masses in the helium shell originate from the star's initial $^{28}$Si and $^{32}$S masses. A simple prescription for non-solar metallicities retains proposition (a), but corrects proposition (b) to

$$^{29}\text{Si}_{\text{shell}} = 1.40 \ \left(\frac{^{29}\text{Si}}{^{28}\text{Si}}\right)_\odot \ ^{28}\text{Si}_{\text{in}} \qquad ^{30}\text{Si}_{\text{shell}} = 1.87 \ \left(\frac{^{30}\text{Si}}{^{28}\text{Si}}\right)_\odot \ ^{28}\text{Si}_{\text{in}} \ . \tag{6}$$

This makes the normalized excesses in the helium shell depend linearly on the initial $^{28}$Si mass. These shell enhancements are mixed and diluted with the envelope, which possess the initial silicon isotope mass fractions. Incorporating proposition (a) gives the normalized excesses as

$$^{29}\text{Si}_{\text{out}} = \left[0.9 + 0.14 \left(\frac{^{29}\text{Si}}{^{28}\text{Si}}\right)_\odot \left(\frac{^{28}\text{Si}}{^{29}\text{Si}}\right)_{\text{in}}\right] \ ^{29}\text{Si}_{\text{in}}$$

$$^{30}\text{Si}_{\text{out}} = \left[0.9 + 0.187 \left(\frac{^{30}\text{Si}}{^{28}\text{Si}}\right)_\odot \left(\frac{^{28}\text{Si}}{^{30}\text{Si}}\right)_{\text{in}}\right] \ ^{30}\text{Si}_{\text{in}} \ . \tag{7}$$

Note that for a solar initial composition, eq. (7) reduces to eq. (5), as it should. Eq. (7) was adopted for the silicon isotope evolutions to be discussed in §3.

Gallino et al. (1994) confirm the assertion that $^{29,30}$Si masses in the helium shell are independent of the initial $^{29,30}$Si masses (see their Table 2), and depend linearly on the initial $^{28}$Si mass. Gallino et al. show that in the helium shell $\delta_\odot^{29}$Si $\sim 400$ and $\delta_\odot^{30}$Si $\sim 900$. Once these excess are mixed with the rest of the AGB envelope, diluting the excess by roughly a factor of 10, they are the same as the factors 1.04 and 1.087 given in eq. (5). It is of relevance in this regard that Gallino et al. give values $\delta_\odot^{29}$Si $= 10$ and $\delta_\odot^{30}$Si $= 23$ in the carbon star phase, where they estimate a shell to envelope ratio of 1/40. These are about a factor of 4 smaller than the values given by eq. (5). However, a carbon star still has a way to go before becoming a planetary nebula. The $^{29,30}$Si s-process production factors given by Gallino et al. and Brown & Clayton (1992) coincide with the simple estimates of this paper. Caution is advised, however. Agreement between the calculations may not be the solution nature chooses. The AGB star may lose a significant fraction of its envelope prior to becoming a carbon star, in which case the $^{29,30}$Si excess factors of eq. (5) are too small. The excess factors are sensitive not only to mass loss, but how many dredgeup episodes occur after the AGB becomes a carbon star. These processes are sufficiently unknown so that the real amplitude of the excess factors is uncertain by perhaps a factor of 10.

## 2.4   INTERLUDE

A very noteworthy situation has arisen. Each of the sources (Type II, Type Ia, and AGBs) makes less $^{29}$Si than $^{30}$Si, yet solar $^{29}$Si is larger than solar $^{30}$Si. *Any* chemical evolution calculation of the silicon isotopes which uses instantaneous mixing, and the three sources employed here, will miss the correct solar $^{29}$Si/$^{30}$Si mass fraction ratio by being about a factor of 3/2 smaller (see §3). This discrepancy must be addressed for the fine details (e.g., parts per thousand deviations) of silicon isotope evolution. Where is the extra $^{29}$Si made in nature?

There are at least four answers to this question, which we state here and discuss below. The first is that some unknown type(s) of star(s) provide an additional source of $^{29}$Si. The second is that the Sun is enriched in $^{29}$Si, being atypical of the mean ISM. The third is that treatment of convection in the 1D stellar models gives an incorrect $^{29}$Si/$^{30}$Si production ratio when averaged over an initial mass function. The fourth is that the details of the nuclear cross sections are modestly in error, so that supernovae produce a $^{29}$Si/$^{30}$Si ratio that is smaller than the solar ratio by roughly a factor of 3/2.

The first alternative seems implausible. An unaccounted source would have to be very prolific, approximately producing half of the galactic content of $^{29}$Si without appreciable $^{30}$Si. Overlooking a source of that magnitude does not seem likely. The second possibility suffers the same weakness; almost half the solar $^{29}$Si would have to have been admixed into it from a nearby source very rich in $^{29}$Si. The third potential answer has merit. As discussed above, 2D hydrodynamic models of convective oxygen shell-burning find plume structures in the velocity field and significant mixing beyond the boundaries defined by mixing-length theory. Although no results have been published yet, it is likely the yields from 2D calculations will differ from the yields from 1D calculations for individual supernova. While integration over an initial mass function smoothes out stochastic yields from stars of different mass or even different yields from the same progenitor mass (e.g., Arnett 1996), it cannot be dismissed that the 2D yields will show enhancements in $^{29}$Si/$^{30}$Si over the 1D models. The fourth alternative also has merit. The nuclear data are inexact, and errors of even tenfold in some key charged particle cross sections (as opposed to neutron capture cross sections) that effect the silicon isotopes are not out of the question. A future study might reexamine the yield dependence on specific reaction rates, the status of the nuclear data upon which the rates are based, by how much the relevant rates might need to be changed, and weather the implied rate changes are within the experimental uncertainties of the present reaction rate. This is beyond the scope of the present paper, and we simply note that changes to the nuclear reaction rates may be the most appropriate answer.

With this noteworthy point in mind, attention is turned to the Galactic evolution of the silicon isotopes and the renormalization of them such that they pass exactly through solar.

## 3. EVOLUTION OF THE SILICON ISOTOPES

The time evolution of the silicon isotopes in the solar neighborhood, culminating in the material from which the Sun was born, and presumably recorded in meteoritic grains, has three principal sources (Type II supernova, Type Ia supernova, and AGB stars). Our treatment of the evolution, based on Timmes et al. (1995), seems reasonably complete; a numerical chemical evolution calculation that incorporates all the detailed nucleosynthetic yields from the massive star survey of Woosley & Weaver (1995), standard paradigm Type Ia supernovae, and estimates of the yields from low mass stars. Consider first the case of homogeneous chemical evolution, in which the ISM at the solar radius has at any time a uniform composition.

Evolution of the silicon isotopes on a traditional stellar abundance ratio diagram is shown in Figure 4. The small inset plot shows the evolution over the entire metallicity range, while the main plot expands the region -1.0 dex $\leq$ [Fe/H] $\leq$ 0.1 dex. [1] A few comments about the global properties of Fig. 4 are in order. Summation of the silicon isotopes, which is dominated by $^{28}$Si, gives the elemental silicon history. Elemental silicon displays many of the trends typical of [$\alpha$-chain nuclei/Fe] ratios – a factor of $\sim$ 3 enhancement in the halo, small mass and metallicity variations, and a smooth drop down to the solar ratio. The departure from classical $\alpha$ element behavior at [Fe/H] $\lesssim$ -2.5 dex in the inset figure is primarily due to uncertainties in the extremely low metallicity 30 M$_\odot$ (and larger) exploded massive star models. However, the general trends of elemental silicon implied by the inset figure are consistent with all known stellar abundance determinations (see Timmes et al. 1995 for details).

Type II supernovae are the principle source for all of the silicon isotopes, with Type Ia supernovae and intermediate-low mass stars making small perturbations. The mean ISM [$^{28}$Si/Fe] ratio in the main plot of Fig. 4 is fairly constant with metallicity, whereas [$^{29}$Si/Fe] and [$^{30}$Si/Fe] increase as time progresses. This is because production of $^{28}$Si by Type II supernovae is primary, being generally independent of the initial metallicity, whereas the $^{29}$Si and $^{30}$Si yields from Type II supernovae are secondary, with their production dependent on the initial metallicity (see Fig. 1). Stars at earlier epochs from a well-mixed ISM have smaller metallicities and smaller secondary/primary ratios. The evolution of the silicon isotopes shown in Fig. 4 is quite different from the one presented in Gallino et al. (1994). The difference is traceable to their assumption or interpretation that the neutron-rich silicon isotope yields are primary instead of secondary (compare their Fig. 1).

Injection rates of the silicon isotopes as a function of time are shown in Figure 5. The age of the Galaxy is taken to be 15 Gyr and the age of the Sun to be 4.5 Gyr. To elucidate the magnitude and direction of the changes induced by each source (Type II, Type Ia, and AGBs), four separate calculations were done. First we describe the procedure used when any of the three sources are added or subtracted, then we describe why this procedure may be optimal, and finally we present an analysis of the figure.

Solid curves in Fig. 5 show the case when all three sources are contributing to $^{28,29,30}$Si. This is the only unambiguous case, and is unaffected by any addition or subtraction procedure. Dotted curves show the evolution when Type II supernovae and AGB stars contribute to changes in $^{29,30}$Si, or equivalently, when Type Ia supernovae contributions to $^{29,30}$Si are removed from the total. The W7 Type Ia masses of $^{29,30}$Si were added into the $^{28}$Si, but all other W7 ejecta (e.g., $^{56}$Fe) contribute in their usual manner. Short dashed curves are for when only Type II and Type Ia supernovae contribute to changes in $^{29,30}$Si, or equivalently, when AGB influences on $^{29,30}$Si are removed from the total. The mass fractions of $^{28,29,30}$Si ejected by AGB stars were set equal to the mass fractions of $^{28,29,30}$Si when the AGB stars were born, but all other AGB ejecta (e.g., $^{12,13}$C) contribute as before. Long dashed curves show the evolution when only Type II supernovae contribute to changes in $^{29,30}$Si, or equivalently, when Type Ia supernovae and AGB stars are removed from the total.

---

[1] The usual spectroscopic notation, [X] = $\log_{10}$(X) - $\log_{10}$X$_\odot$ for any abundance quantity or ratio X, is adopted.

It is extremely difficult to extract meaningful statements under the seemingly "straightforward" approach of starting with only Type II supernovae, adding in Type Ia supernovae, adding in AGB stars, and then examining the sum of all three. First, the elemental silicon curves Si(t) for each case will not be the same. Each elemental silicon history takes a different amounts of time to reach a given [Fe/H]. If Type Ia events are naively removed, then important iron contributions are removed, and metallicity based chronometers become unsynchronized. Second, the isotopic composition at distances and times appropriate for the presolar nebula are different as each source is activated. Each case will not produce an isotopic solar composition at the level attained in Figure 5 of Timmes et al. (1995). There are also ancillary issues of star formation rates and present epoch supernova rates becoming unacceptably large or small as various sources are added or removed. Thus, it is hard to interpret abundance trends under the seemingly "straightforward" approach and may even be inconsistent. On the other hand, the procedures described above for subtracting the $^{29,30}$Si contributions from a source assures that elemental silicon evolves in *exactly* the same manner in each case. All of the sources occur in nature, and one does not want to "turn off the source". We want to know how important the $^{29,30}$Si contributions of a particular source are, so we adjust the yields so as to produce the identical elemental silicon evolutions Si(t). An unchanging elemental silicon evolution allows a sharper delineation of changes in the silicon isotopic composition induced by each source. Any changes in the isotopic ratios are due to changes in $^{29}$Si and $^{30}$Si, not to changes in $^{28}$Si.

Since the mass of $^{28}$Si returned to the ISM is the same in each calculation, all four $^{28}$Si curves overlie each other in Fig. 5. By comparing the two curves (short-dash and long-dash) which exclude Type Ia supernovae contributions to $^{29}$Si with the two curves (dotted and solid) that include them, we conclude that the effect on the injection rates of $^{29}$Si when Type Ia supernova are added or removed from the mixture is negligible. An order of magnitude more $^{30}$Si is ejected than $^{29}$Si by the W7 model, and is the only isotope shown in Fig. 5 that crisply separates the various contributions. Exclusion of Type Ia contributions to $^{30}$Si (dotted curve) reduces the $^{30}$Si injection rate by a few percent, while exclusion of AGB contributions to $^{30}$Si (short-dash curve and eq. 8) gives a slightly smaller injection rate. Removal of Type Ia and AGB contributions to $^{30}$Si (long-dash curve) reveals the dominance of core collapse events in the injection of $^{30}$Si into the ISM.

Although Type II supernovae are chiefly responsible for setting the absolute abundance levels and the injection rates of the silicon isotopes into the ISM, both AGB stars and Type Ia supernova add discernible perturbations. The return fraction from AGB stars begins small, due to their longer lifetimes, but grows larger as time increases. At the time the Sun formed our analysis suggests about 75% of the silicon isotopes being ejected was freshly synthesized silicon from massive stars, about 20% was the return of previously synthesized silicon from AGBs (slightly modified by s-processing), and about 5% was new silicon synthesized from Type Ia events. The ejecta of these three sources follow differing adiabats, are exposed to different radiation environments, mixing mechanisms, mixing timescales, and grain formation timescales. Grains which have condensed from a well-mixed mean ISM should, in general, have isotopic compositions reflective of their differing pathways. This is the idea underlying "cosmic chemical memory" in presolar grains from meteorites (Clayton 1982).

Fig. 4 already displays ramifications of the situation discussed in §2.4. The calculated $^{29}$Si/$^{30}$Si ratio is smaller than the solar ratio by roughly 0.2 dex, a of factor 1.5 on linear scales. No possibility exists for this, or any other, homogeneous calculation to reproduce the silicon isotope ratios with the precision necessary for comparison with presolar meteorite grains. To circumvent this, one can renormalize the curves to the calculated silicon isotope composition at solar birth. This is roughly equivalent to changing the all $^{28,29,30}$Si yields from massive stars by 3/2, and may be viewed, per §2.4, as a small systematic correction to the underlying nuclear data base or as a correction due to treating convection more rigorously. Renormalization allows an apples-to-apples comparison of measured SiC silicon isotope ratios with the calculations and concepts of chemical evolution. It is self-consistent in that experimental data are compared with the composition that supernovae themselves produce, not with a composition that supernovae do not produce. Differences between normalization with the calculated ISM silicon isotopic composition and solar composition is a central concept of this paper.

An example of this renormalization procedure is the evolution of the silicon isotopes in a three-isotope plot shown in Figure 6. The variational procedure and meaning of the various curve types (solid, dotted, short-dash, and long-dash) are the same as discussed for Fig. 5. Deviation of the silicon isotopes from their values calculated at a place (8.5 Kpc Galactocentric radius) and time (10.5 Gyr in a 15 Gyr old Galaxy) appropriate for the presolar nebula were used for the axes (note subscript) and the curves. That is, deviations are expressed not with respect to solar, which the calculation does not pass through, but with respect to the values calculated at solar birth.

The normalizing silicon isotope mass fractions, when all three sources of silicon are contributing, were taken to be

$$X(^{28}Si)_{ISM} = 9.70 \times 10^{-4}$$
$$X(^{29}Si)_{ISM} = 4.77 \times 10^{-5} \quad (8)$$
$$X(^{30}Si)_{ISM} = 5.09 \times 10^{-5} \ .$$

This is quite similar to the silicon isotopic composition shown in Fig. 5 of Timmes et al. (1995), the difference attributable to $^{29,30}$Si enhancements from AGB stars (eq. 7). For comparison, the Anders & Grevesse (1989) silicon isotope mass fractions are

$$X(^{28}Si)_\odot = 6.53 \times 10^{-4}$$
$$X(^{29}Si)_\odot = 3.43 \times 10^{-5} \quad (9)$$
$$X(^{30}Si)_\odot = 2.35 \times 10^{-5} \ .$$

These two compositions have different isotopic ratios because of the noteworthy situation discussed in §2.4, namely; each source makes less $^{29}$Si than $^{30}$Si, and yet solar $^{29}$Si is larger than solar $^{30}$Si. Relative to the solar $X(^{29}Si)/X(^{30}Si)$ ratio, the normalizing ISM composition has a ratio that is a factor of $1.557 \simeq 3/2$ smaller, as alluded to in §2.4. Bulk supernova ejecta when normalized by the mean ISM silicon isotopic composition of eq. (8) are given in the middle two columns of Table 1, and in the last two columns of Table 1 when normalized by the solar composition of eq. (9). Any chemical evolution calculation of the silicon isotopes which uses instantaneous mixing, and the three sources used here, will be smaller than the correct solar $^{29}$Si/$^{30}$Si mass fraction ratio by roughly a factor of 3/2. Renormalization causes deviations to pass exactly through the origin at 10.5 Gyr. Other ages for the galaxy simply rescale the time values shown in Fig. 6.

The isotopic evolution marches up the solid line at a rate measured by the time arrows on the right in Fig. 6. The correlation line has slope near unity, m=0.975 for the solid line, in agreement with Clayton (1988). As anticipated in §2.1 from Fig. 2, mean chemical evolution models, whose nucleosynthesis is dominated by ejecta from core collapse events, produce m=1 slope lines in a three-isotope plot *when* the mean evolutions are normalized with respect to the calculated silicon isotopic composition at solar birth (eq. 9). It would not be unity slope line if the solar normalization (eq. 8) were used. This crucial point is analyzed in detail and explicitly demonstrated in §4.6.

The largest slope in Fig. 6 occurs when only Type II events contribute (long-dash line) to $^{29,30}$Si. Type Ia supernova and AGB stars make small perturbations (short-dash and dotted lines) compared to the net result (solid line) when all three sources contribute to $^{29,30}$Si. The small effect of AGB stars, even with the generous prescription of eq. (7), confirms that any coefficient errors in eq. (7) are unimportant for mean chemical evolution (though of importance for AGB stars themselves). Fig. 6 strongly suggests that chemical evolution models which employ instantaneous mixing of stellar ejecta into the bulk ISM cannot produce slopes much different than unity.

Silicon isotopic compositions of Murchison SiC samples measured by Hoppe et al. (1994) have a best-fit slope of 4/3 and are shown in Fig. 6. The grains are located by their deviations with respect to solar, whereas the chemical evolution curves are located by deviations with respect to the calculated silicon isotopic composition at solar birth. These two representations are equal, $\delta_\odot = \delta_{\rm ISM}$, in the renormalization picture. Most of the mainstream grains shown in Fig. 6 have a positive $\delta^{29}$Si and $\delta^{30}$Si. If this trend is attributed to a mean ISM, this requires AGB stars that formed later than the Sun. Clearly, an AGB star born after the Sun could not have mixed its SiC grains into the presolar cloud. Inhomogeneous pockets that are later mixed with the mean ISM (Malinie et al. 1993) could give a presolar nebula that has a negative $\delta^{29}$Si and $\delta^{30}$Si with respect to the mean ISM at that time. In addition, several studies have revealed a spread in the atmospheric abundances of dwarf stars at any given metallicity or age (e.g., Wheeler et al. 1989; Edvardsson et al. 1992), indicating that some evolutionary effects involve the incomplete mixing of stellar ejecta with the ISM. As such, signatures from inhomogeneous mixing is a subject to which we now turn.

## 4. DIFFERING ISOTOPIC RESERVOIRS AND SiC GRAINS

For isotopically anomalous SiC grains to exist requires at least two conditions. Firstly, nature must provide distinct isotopic pools from which they may be grown. Secondly, nature must provide a machine for manufacturing the SiC grains from those pools of matter. The problem is to identify both the pools and the machine. Several interpretations of both are now explored.

## 4.1 RECENT STARDUST IN BULK

The simplest case of differing isotopic pools is recent ejecta and bulk ISM. If condensates from cooling stellar ejecta are rapidly destroyed by sputtering (primarily), melting and vaporization processes in the ISM, then any grains which exist today must be young and must have condensed out of recent ejecta. Clayton (1988) calculated that $^{29,30}$Si would be $\sim 56\%$ (the numerical evolution gives 59%) more abundant than $^{29,30}$Si in the ISM at solar birth (i.e., grains which condense from this material are enriched in both secondary isotopes by 59%). Young condensates are too simple an explanation of SiC grains, however, for at least three reasons: (1) the correlation slope is not the measured 4/3 value of mainstream grains; (2) the SiC grains carry s-process signatures (Lewis et al. 1994; Ott & Begemann 1990; Prombo et al. 1993), although it cannot be stated that all SiC grains carry it; (3) the carbon isotopic compositions in SiC grains vary greatly in uncorrelated ways, whereas bulk ejecta is simply $^{13}$C enriched. Young condensates cannot be the SiC machines; SiC grain compositions constrain and select carbon-rich layers from stars as sources.

## 4.2 GASEOUS STELLAR EJECTA AND OLD GRAINS

Suppose all stellar ejecta is gaseous. Grain mass and composition are then set by gaseous accretion onto preexisting nucleation sites. Under these conditions, the smallest grains will be the most enriched in freshly ejected $^{29,30}$Si (Clayton 1980; Clayton, Scowen & Liffman 1989). Although this picture may work for some of the correlated $^{48}$Ca, $^{50}$Ti, $^{54}$Cr, $^{58}$Fe, $^{64}$Ni, and $^{66}$Zn excesses in solar system solids (Clayton 1981), it fails as an explanation for presolar SiC grains for the same objections given above. Exceptions could occur if it is chemically possible to preferentially accrete gaseous silicon and carbon, although there is no evidence from material sciences that SiC can be grown from anything but a carbon-rich gas at high temperatures. In addition, accretion of isotopically homogeneous dust still puts silicon isotopic ratios on a m=1 line, not a m=4/3 line.

## 4.3 STARDUST FROM STARS OF DIFFERING AGES

Suppose the STARDUST machines are stars that formed at different epochs. Since $^{29,30}$Si increase monotonically (Fig. 4), one can use their abundance levels as a chronometer. Under these conditions, a sequence of points in a three-isotope plot may be interpreted as a chronological sequence, with different ages for different grains. If grains inherit an isotopic composition equal to the initial composition of the star, the oldest grains will be the most deficient in the secondary isotopes (Clayton, Scowen & Liffman 1985). This mechanism works most simply for Wolf-Rayet stars, which evolve on such a rapid timescale that their return is approximately instantaneous. This time correlation picture is not so direct for AGB stars since different stellar masses have different lifetimes, which introduces a dispersion in silicon compositions that is difficult to disentangle.

## 4.4 SPECIFIC NUCLEAR EFFECTS

The correlations shown in Fig. 6 are remarkably robust with respect to variations in the initial mass function, stellar birth rate, infall time scales, and assumed ages for the Galaxy. Evidently chemical evolution models which employ instantaneous mixing of stellar ejecta into the bulk ISM cannot produce slopes much different than unity. Thus, homogeneous chemical evolution by itself cannot completely explain the anomalous silicon isotope ratios in presolar SiC grains. A complete solution requires an anomalous isotopic pool that does not lie on the slope m=1 line. That anomalous pool might be within the stars themselves, for anomalous pools certainly exist within stellar interiors, or the inhomogeneous contamination of the material from which the stars formed. Either pool might cooperate with homogeneous chemical evolution to produce the correlation measured in SiC grains, and an example involving a hypothesized metallicity trend in AGB stars follows.

### 4.5 AN EXAMPLE AGB CORRELATION LINE

AGB stars of differing metallicity may be the machines which make the SiC grain distribution, an idea that has been discussed extensively (e.g., Gallino et al. 1994). Consider two AGB stars on the m=1 slope line, each with a different initial metallicity, hence different silicon isotope, as shown in Figure 7. Since silicon isotopic ratios in the helium shell after thermal pulsations are independent of the initial silicon isotopic composition (see §2.3), both AGB star's helium shell silicon isotope compositions map to a single point in a three-isotope plot. This unique shell composition is labeled as "S" in Figure 7. For clarity, Fig. 7 is drawn as a schematic rather than to scale, but this doesn't change the qualitative features which follow.

During dredgeups the envelopes of these AGB stars are mixed with shell matter, with the mixed composition being a linear combination of the initial envelope composition and the unique shell composition S. Mixtures of two compositions in a three-isotope plot must, mathematically, lie along the line connecting the two end points. Furthermore, the relative numbers of nuclei contributed by each point is inversely proportional to the distance between the mixtures and the point. The situation is like weights balanced on a lever, with the mixed composition being the fulcrum. Thus, mixtures between the AGB envelopes and the shell composition must lie along the lines drawn between the two AGB stars and the point S in Fig. 7.

Now let Sf1 and Sf2 represent the fraction of shell material mixed with the envelope in each star at the time when SiC grains form and depart. Sf1 and Sf2 are $\simeq 10\%$ during the carbon star phase, but may be larger in later phases when the strongest winds eject the greatest density of atoms for SiC nucleation. If Sf1 and Sf2 are equal in stars of different initial metallicity, both mix points (labeled "1" and "2") will be shifted by the same degree towards S. In this case, the SiC grains still correlate along a m=1 line, but shifted to the right of the original m=1 line.

Consider the hypothetical case of the lower metallicity star (AGB1) having a larger fraction of shell material mixed into its envelope than the higher metallicity star (AGB2). That is, let Sf1 > Sf2. Then it is easy to see that point "1" is moved farther to the right than point "2" in Fig. 7. The line connecting the two mix points now has slope steeper than unity. Under the right conditions, it may have the measured m=4/3 slope. In addition, if the degree of shell and envelope mixing is linear with metallicity, then all SiC grains correlate along the m >1 line. Furthermore, should the m=1 line pass through the solar isotopic composition, the m > 1 line will pass to the right of the solar composition.

A quantitative estimate for how much larger a fraction of shell material needs to be mixed under this scenario is useful. Let AGB2 have a solar silicon composition, $\delta^{29}_{\text{AGB2}} = \delta^{30}_{\text{AGB2}} = 0$. The unique silicon isotopic shell composition S, which enriches solar $^{29}\text{Si}/^{28}\text{Si}$ ratios by 40% and solar $^{30}\text{Si}/^{28}\text{Si}$ ratios by 87% (see §2.3), has $\delta^{29}_{\text{S}} = 400$, $\delta^{30}_{\text{S}} = 870$. Let the shell mixing fraction Sf2 of AGB2 be the canonical 10% when it's SiC grains form. The silicon composition of this mixed material is $\delta^{29}_{\text{mix2}} = 40$, $\delta^{30}_{\text{mix2}} = 87$. Now place AGB1 on the m=1 slope line by assigning it to have the arbitrary values $\delta^{29}_{\text{AGB1}} = \delta^{30}_{\text{AGB1}}$. A 4/3 slope between mix point 2 and mix point 1 requires

$$\frac{4}{3} = \frac{\delta^{29}_{\text{mix2}} - \delta^{29}_{\text{mix1}}}{\delta^{30}_{\text{mix2}} - \delta^{30}_{\text{mix1}}} = \frac{\delta^{29}_{\text{mix2}} - ((1-\text{Sf1}) \cdot \delta^{29}_{\text{AGB1}} + \text{Sf1} \cdot \delta^{29}_{\text{S}})}{\delta^{30}_{\text{mix2}} - ((1-\text{Sf1}) \cdot \delta^{30}_{\text{AGB1}} + \text{Sf1} \cdot \delta^{30}_{\text{S}})} \quad . \tag{10}$$

Solving for the shell mixing fraction Sf1 of AGB1 gives

$$\text{Sf1} = \frac{3\delta^{29}_{\text{mix2}} - 4\delta^{30}_{\text{mix2}} + \delta_{\text{AGB1}}}{3\delta^{29}_{\text{S}} - 4\delta^{30}_{\text{S}} + \delta_{\text{AGB1}}} = \frac{\delta_{\text{AGB1}} - 228}{\delta_{\text{AGB1}} - 2280} \quad . \tag{11}$$

For the case $\delta^{29}_{\text{AGB1}} = \delta^{30}_{\text{AGB1}} = $ -260, the shell mixing fraction Sf1 of AGB1 is 19%, roughly twice as large as the shell mixing fraction Sf2 of AGB2. Mix point 1 then has $\delta^{29}_{\text{mix1}} = $ -133, $\delta^{30}_{\text{mix1}} = $ -43.

One could object that we have merely postulated an effect that will achieve the desired result. That is correct, but our hypothetical case is not implausible either. For example, the wind strength in most mass loss formulations depends upon the initial metallicity. The lower metallicity star has a weaker wind and thus sustains more shell flashes and dredgeups during its lifetime before the overlying envelope mass becomes inadequate. With more dredgeup episodes, a lower metallicity star may have a larger envelope-shell mixing fraction than a higher metallicity star. Detailed stellar models and isotopic abundance determinations from AGB star observations are the final arbitrator of this hypothetical mechanism.

### 4.6 INHOMOGENEOUS ENRICHMENT OF STAR FORMING REGIONS

Inhomogeneous enrichment of star forming regions is a mechanism to produce metallicities distinct from the mean ISM. If formation of a suite of AGBs whose initial silicon isotopic compositions correlate along a slope 4/3 line were instigated by a single specific supernova that formed earlier in the same association, then in one-stage enrichment scenarios such as this one, the supernova ejecta would have to be displaced from the initial silicon isotopic composition along a 4/3 slope line. With two-stage enrichment scenarios, more pathways exist and obtaining a well-defined correlation line from multiple physical histories is more difficult. It is thus useful to examine one particular set of massive stars and the composition into which their supernovae ejecta is mixed.

Type II silicon ejecta mixed with either the computed silicon isotopic composition at the time of solar birth or with the Anders & Grevesse (1989) solar composition is shown in Figures 8a, 8b, and 8c. Magnitudes of the vectors in nature are determined by how much of a supernova's ejecta is mixed in with the ambient medium. Fig. 8 used a mix fraction of 0.001; i.e., 1 g of supernova ejecta uniformly mixed with 1 kg of ambient material. Other dilution factors scale the vector lengths proportionately. Any linear mixture of the ambient material with the ejecta must lie along the source's vector.

Fig. 8a shows the massive star yields mixed with the computed ISM (eq. 9) at the time of solar birth. Deviations in Fig. 8a are expressed with respect to this mean ISM (note coordinate subscripts), rather than with respect to solar, and the dotted m=1 slope line is the mean chemical evolution line of Fig. 6. This renormalization admits the interpretation that this is a shift of the calculated ISM by the calculated admixtures (see §2.4). If the solar composition is made to fall on the mean ISM evolution by renormalization, as is done here, then and only then does ISM mean solar, otherwise they are not the same. Fig. 8a represents a self-consistent chemical evolution when referenced by a system lying on that mean evolution. Bulk supernova ejecta, diluted by the computed ISM composition and solar, are listed in Table 2 for both normalization choices. The middle two columns of Table 2 are the end points of the vectors shown in Fig 8a. Note that all of the mixing vectors point within a small opening angle of the m=1 correlation line; none of the mixing vectors make a 90° angle to the mean evolution line. As the mass of mean ISM mixed with the supernova mass yields is increased (dilution factor increased), the length of the mixing vectors decreases toward the proper $\delta^{29}_{\rm ISM}=0=\delta^{30}_{\rm ISM}$ reference point. Note that the envelope of the mixing vectors in Fig. 8a possesses the same shape as the mainstream SiC grains shown in Fig. 6.

Massive star yields mixed with the computed ISM at the time of solar birth (eq. 9), but with deviations expressed with respect to solar (eq. 10) are shown in Fig. 8b. This case illuminates the differences between normalization bases from which deviations are evaluated. This figure regards the calculated evolution as being the correct mean evolution, but viewed from a third system (the solar system) that does not lie on that mean evolution. The points shown in Fig. 8b are the same points as in Fig. 8a, only the reference frame has changed. These two reference bases are connected by the simple linear coordinate transformation

$$\delta^{29}_{\odot} = 1000 \left[ \frac{^{29}{\rm Si}/^{28}{\rm Si}}{(^{29}{\rm Si}/^{28}{\rm Si})_{\odot}} - 1 \right] = 1000 \left[ \frac{^{29}{\rm Si}/^{28}{\rm Si}}{(^{29}{\rm Si}/^{28}{\rm Si})_{\rm ISM}} \cdot \frac{(^{29}{\rm Si}/^{28}{\rm Si})_{\rm ISM}}{(^{29}{\rm Si}/^{28}{\rm Si})_{\odot}} - 1 \right]$$
$$= 1000 \left[ \frac{(^{29}{\rm Si}/^{28}{\rm Si})_{\rm ISM}}{(^{29}{\rm Si}/^{28}{\rm Si})_{\odot}} \cdot \left( \frac{\delta^{29}_{\rm ISM}}{1000} + 1 \right) - 1 \right] \quad , \tag{12}$$

and similarly for $\delta^{30}_{\odot}$. Substituting the values given in eqs. (9) and (10), gives the simple expressions:

$$\delta^{29}_{\odot} = 0.937 \, \delta^{29}_{\rm ISM} - 63 \qquad \delta^{30}_{\odot} = 1.458 \, \delta^{30}_{\rm ISM} + 458 \quad . \tag{13}$$

This expresses a translation and a rotation in three-isotope diagrams. As a result, the m=1 line of Fig. 8a is rotated into the m=2/3 line of Fig. 8b. Relative to solar, the calculated mean ISM silicon composition is $^{29}$Si poor and $^{30}$Si rich. The mean ISM is shifted from $\delta^{29}_{\rm ISM} = \delta^{30}_{\rm ISM} = 0$ in Fig. 8a to $\delta^{29}_{\odot} = -63, \delta^{30}_{\odot} = 458$ in Fig. 8b. As the dilution factor is increased the length of the mixing vectors decreases toward the $\delta^{29}_{\odot}$ = -63, $\delta^{30}_{\odot} = 458$ origin. This shift is emphasized in Fig. 8b by the arrow pointing towards the origin of a solar silicon reference frame. Mixing vectors in Fig. 8b point in different in directions, with small amplitude changes, despite being the same data as Fig. 8a. This occurs because the supernova yields do not produce a chemical evolution that passes exactly through the solar silicon point. Normalizing with the silicon isotopic composition at solar birth make the evolution pass exactly through the solar point, and in so doing Fig. 8b would become identical to Fig. 8a. For quantitative considerations, the last two columns of Table 2 are the end points of the vectors shown in Fig 8b.

Supernovae between 30—40 $M_\odot$ produce quite different correlation slopes, as seen by their different vector directions in Figs. 8a and 8b. The directional differences are due to the larger fallback mass in the more massive stars. A significantly larger fraction of $^{28}$Si falls back onto the compact remnant since it is synthesized closer to a star's center than the heavier silicon isotopes. While the total mass that experiences fallback in the stellar models is uncertain, it is not physically unreasonable, but it is probably only a lower limit since matter accreted during the first second of the delayed explosion mechanism is neglected. For the case of Fig. 8b, slightly more massive stars are required to produce a m=4/3 correlation slope than in the representation of Fig. 8a.

Fig. 8c shows the case when massive star yields are mixed with solar abundances and deviations are expressed with respect to the solar. This case has the interpretation that the Sun formed from a solar silicon cloud complex, even though the supernova yields do not generate exactly such a mean silicon composition. Surprisingly, the innocent act of combining solar metallicity massive star yields with deviations expressed with respect to solar abundances is not self-consistent, but it is often discussed in relationship to SiC and graphite grains. Mixing vectors in this reference frame point in directions that only appear to be unpromising for generating a 4/3 slope line, when in fact they are quite promising when a proper reference frame (Fig. 8a) is established. The unpromising quandary arises in the first place because deviations expressed with respect to solar is inconsistent with the composition that massive stars produce. The final two columns of Table 2 list the endpoints of the mixing vectors shown in Fig. 8c.

The inhomogeneous mixture scenario represented by Fig. 8a seems the most plausible for generating a 4/3 correlation slope. It takes the mean ISM to have a solar silicon composition and dilutes it differentially with various supernova ejecta. This may spawn many correlated stars. Even in this favorable case it can be difficult to imagine how the secondary stars, those AGB machines which manufacture SiC, so easily emulate a 4/3 correlation among their initial compositions. It could be, or could not be, as simple as having the slopes of gas enriched by high-mass supernovae (30–40 $M_\odot$) and the slopes enriched by less massive supernovae average to a mean 4/3 slope.

### 4.7 SILICON ISOTOPES IN THE X GRAINS

The introduction described a class of SiC grains from meteorites, the X grains, that appear to be supernova condensates (SUNOCONs) based on the specific non-silicon isotopic signatures that they carry. The silicon isotopic patterns in these grains have been difficult to understand since the bulk $^{29,30}$Si supernova yields appears not to be compatible with the strong $^{28}$Si richness of these grains (Amari et al. 1992; Nittler et al. 1995ab; Hoppe et al. 1996). Our suggested solution to the impasse presented by the mainstream grains is a renormalization such that chemical evolutions pass exactly through the solar silicon isotopic composition. This renormalization may also help with the problem presented by SiC X grains. To test this quantitatively, Figure 9 shows the locations of the known X-type SiC grains with the undiluted and ISM normalized yields of Table 1. Silicon isotopic compositions of Murchison SiC samples measured by Hoppe et al. (1996; unpublished) and Nittler et al. (1995ab) are located by deviations with respect to solar isotopic abundances $\delta_\odot$, whereas the undiluted supernova ejecta are located by deviations with respect to the mean ISM at solar birth $\delta_{\rm ISM}$. These two are the same $\delta_\odot = \delta_{\rm ISM}$ under renormalization (Fig. 8a). This figure suggests that X-type SiC grains have silicon isotopic compositions that one would expect from the bulk ejecta of the most common Type II supernovae.

Fig. 8c illustrates the difficulty X grains present when viewed from calculation that is inconsistent. Most of the mixing vectors from common solar metallicity supernovae appear too deficient in $^{29}$Si. to explain the X grains, which contain a $^{29}$Si/$^{30}$Si ratio greater than solar, but diluted with an excess $^{28}$Si. Supernovae, especially those with a smaller mass, seem much more promising sources in a self-consistent renormalized-yield calculation (Figs. 8a and 9).

A perhaps astonishing coincidence arises when we view the 30-40 $M_\odot$ supernovae in this regard. If 12–20 $M_\odot$ stars condense X-type SUNOCON SiC, one should expect 30–40 $M_\odot$ stars to do so as well. The more massive progenitors are simply less frequent. A corollary to this line of thought is X-type SiC must exist having $^{29,30}$Si excess as well as deficits, as graphite grains do.

As noted above, the envelope of the mixing vectors in Fig. 8a possesses the same shape as the mainstream SiC grains shown in Fig. 6. If SUNOCON cores could be differentially diluted with the mean ISM, they could produce grains having the same distribution of Fig. 6 and Fig. 9 combined – a line of slope 4/3 (as in the 35 $M_\odot$ mix), a bowing around to the right of the ISM composition, and the $^{28}$Si-rich portion (as in 11–15 $M_\odot$ stars). How this might happen chemically is uncertain and one would also have to account for the wide range of carbon isotopic ratios measured in SiC grains by further processing through AGB stars. In addition, the magnitude of the extinct $^{44}$Ti and $^{49}$V anomalies seem to require that the calcium and titanium in SUNOCON SiC grains were chiefly those calcium and titanium atoms from its initial SUNOCON core. But for all these implausibilies, one might question whether the mainstream SiC represents AGB grains, or whether there is also a healthy mix of diluted SUNOCONs among them. Note supernovae also carry s-process Xe throughout their interiors, anywhere where neutrons have been liberated, so the existence of s-process Xe does not in itself demand AGB origin, although agreement with the krypton data in better with AGB stars than for massive stars.

## 5. SUMMARY

We submit these answers to the questions posed in the abstract:

1. The absolute abundances levels and injection rates of the silicon isotopes into the bulk ISM are dominated by the ejecta of Type II supernovae (Figs. 4 and 5). Almost 80% of $^{28}$Si appears as "new Si" from Type IIs, and even larger percentages hold for $^{29,30}$Si. Type Ia supernova and AGB stars are perturbations on the pattern established by massive stars.
2. The isotope $^{28}$Si is a primary nucleosynthesis product, since its yield is insensitive to the initial metallicity (Fig. 1a), while $^{29,30}$Si are secondary nucleosynthesis products, since their yields depending approximately linearly on the initial metallicity (Figs. 1b and 1c).
3. Mean chemical evolution models produce m=1 correlation slopes in three-isotope diagrams (Figs. 6 and 8a). More massive Type II progenitors move silicon approximately up the m=1 direction, whereas less massive progenitors tend to move it down this correlation line. This difference is due to a larger fallback fraction of $^{28}$Si in the more massive progenitors.

4. The raw evolutions do not pass exactly through the solar isotopic composition. Renormalization with respect to the computed silicon isotopic composition corrects this effect, and offers insights in how deviations are to be viewed (§2.4, Figs. 8a, 8b and 8c). Other trace elements, particularly calcium and titanium, in SiC grains might be addressed by the renormalization procedure.

5. Chemical evolution might have been recorded in SiC grains. Homogeneous m=1 slope evolutions could combine with a metallicity or age effect on the fraction of shell matter mixed with the AGB envelope at the time of SiC condensation to yield a 4/3 correlation line (Fig. 7).

Finally, the silicon isotopic ratios found in X-type SiC grains may be representative of bulk silicon supernova ejecta. This possibility is evident when a self-consistent picture of solar metallicity (Figs. 8a and 9) is used. As a result, we predict that $^{29,30}$Si-rich SiC SUNOCONs will be discovered, just as they have been discovered for graphite grains. The rich data base on SiC grains has opened unique windows in astronomy. This survey may enable a more meaningful assessment of their information content.

The authors thank Ernst Zinner and Peter Hoppe for their unpublished SiC X grain data. We also thank Larry Nittler, Conel Alexander, Roberto Gallino, and Friedel Thielemann for stimulating discussions on the measured silicon isotopes in SiC grains and chemical evolution of the silicon isotopes. Finally, the authors are very grateful for the detailed and thoughtful review of this work by the referee Ernst Zinner.

This work has been supported at Clemson by the W. M. Keck Foundation, by a NASA Planetary Materials and Geochemistry grant (D.D.C), and by a Compton Gamma Ray Observatory Postdoctoral Fellowship (F.X.T); and at Chicago by an Enrico Fermi Postdoctoral Fellowship (F.X.T).

**TABLE 1**

Silicon Isotopic Ratios and Deviations for Solar Metallicity Type II Supernovae Bulk Ejecta[1]

| Mass ($M_\odot$) | $^{29}Si/^{28}Si$ [2] | $^{30}Si/^{28}Si$ [2] | $\delta^{29}_{ISM}$ | $\delta^{30}_{ISM}$ | $\delta^{29}_\odot$ | $\delta^{30}_\odot$ |
|---|---|---|---|---|---|---|
| 11 | 0.0297 | 0.0321 | -396 | -388 | -434 | -108 |
| 12 | 0.0120 | 0.00838 | -755 | -840 | -771 | -767 |
| 13 | 0.0248 | 0.0336 | -496 | -359 | -528 | -66 |
| 15 | 0.0225 | 0.0280 | -543 | -466 | -572 | -222 |
| 18 | 0.0268 | 0.0395 | -455 | -248 | -489 | 97 |
| 19 | 0.0155 | 0.0124 | -685 | -763 | -705 | -655 |
| 20 | 0.0212 | 0.0202 | -569 | -615 | -596 | -439 |
| 22 | 0.0344 | 0.0395 | -300 | -248 | -345 | 97 |
| 25 | 0.0365 | 0.0399 | -257 | -239 | -304 | 109 |
| 30 | 0.107 | 0.109 | 1177 | 1078 | 1040 | 2030 |
| 35 | 0.265 | 0.214 | 4380 | 3074 | 4040 | 4940 |
| 40 | 0.302 | 0.164 | 5144 | 2127 | 4750 | 3560 |

[1] For the Woosley & Weaver (1995) supernovae models.
[2] Isotopic ratios of the bulk ejecta, unnormalized to any composition.

## TABLE 2

Deviations of Solar Metallicity Type II Supernovae Bulk Ejecta with ISM and Solar Dilutions[1]

| Mass ($M_\odot$) | Diluted with ISM[2] | | Diluted with ISM[2] | | Diluted with Solar[2] | |
| --- | --- | --- | --- | --- | --- | --- |
| | $\delta^{29}_{\rm ISM}$ | $\delta^{30}_{\rm ISM}$ | $\delta^{29}_\odot$ | $\delta^{30}_\odot$ | $\delta^{29}_\odot$ | $\delta^{30}_\odot$ |
| 11 | -0.91 | -0.90 | -64.3 | 457 | -1.49 | -0.37 |
| 12 | -6.58 | -7.32 | -69.6 | 448 | -9.93 | -9.88 |
| 13 | -2.55 | -1.85 | -65.8 | 456 | -4.02 | -0.50 |
| 15 | -4.48 | -3.85 | -67.6 | 453 | -6.99 | -2.71 |
| 18 | -3.82 | -2.08 | -67.0 | 455 | -6.08 | 1.20 |
| 19 | -10.9 | -12.1 | -73.6 | 441 | -16.5 | -15.3 |
| 20 | -8.89 | -9.62 | -71.8 | 444 | -13.7 | -10.1 |
| 22 | -5.23 | -4.32 | -68.3 | 452 | -8.84 | 2.48 |
| 25 | -3.47 | -3.23 | -66.7 | 454 | -6.06 | 2.17 |
| 30 | 13.2 | 12.1 | -51.1 | 476 | 17.2 | 33.6 |
| 35 | 14.9 | 10.5 | -49.4 | 474 | 20.4 | 25.0 |
| 40 | 7.45 | 3.08 | -56.5 | 463 | 10.2 | 7.65 |

[1] For the Woosley & Weaver (1995) supernovae models.
[2] Dilution factors are 1000, i.e., 1 g of supernova ejecta mixed with 1 Kg ISM or solar composition.

# Figures and Captions

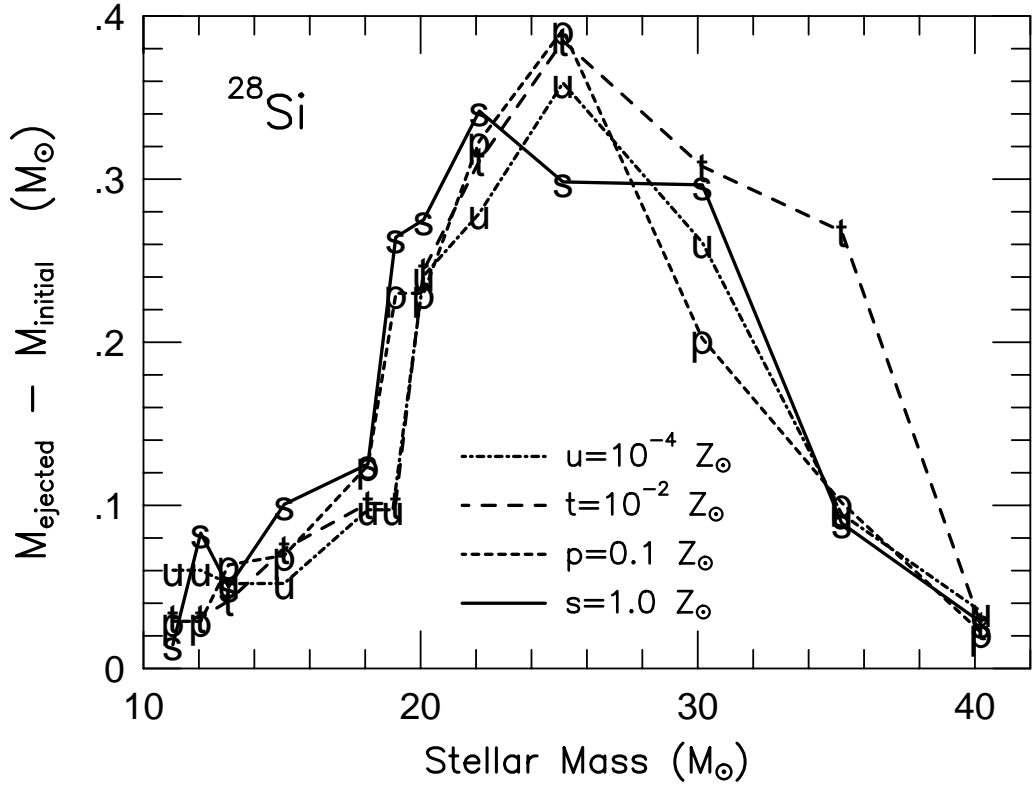

Fig. 1a.— Mass of $^{28}$Si produced from the set of exploded massive star models of Woosley & Weaver (1995). The stellar models whose initial metallicity is $10^{-4}$ Z$_\odot$ are labeled with the letter "u", "t" for the $10^{-2}$ Z$_\odot$ initial metallicity models, "p" for the 0.1 Z$_\odot$ models, and "s" for 1.0 Z$_\odot$. The chief point is that production of $^{28}$Si is primary; the mass ejected is generally independent of the initial metallicity of the star.

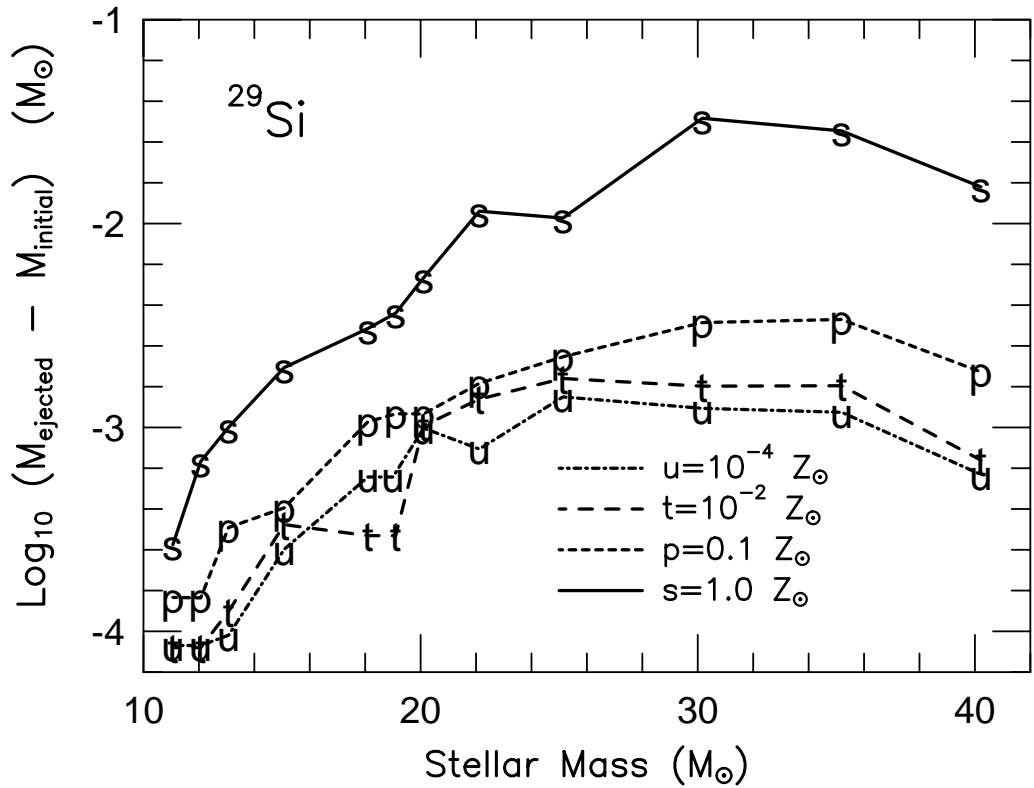

Fig. 1b.— Mass of $^{29}$Si produced from the set of exploded massive star models of Woosley & Weaver (1995). Meaning of the labeled points is the same as in Fig. 1a. The $^{29}$Si yields are very dependent on the initial metallicity of the massive star, hence, its production is secondary.

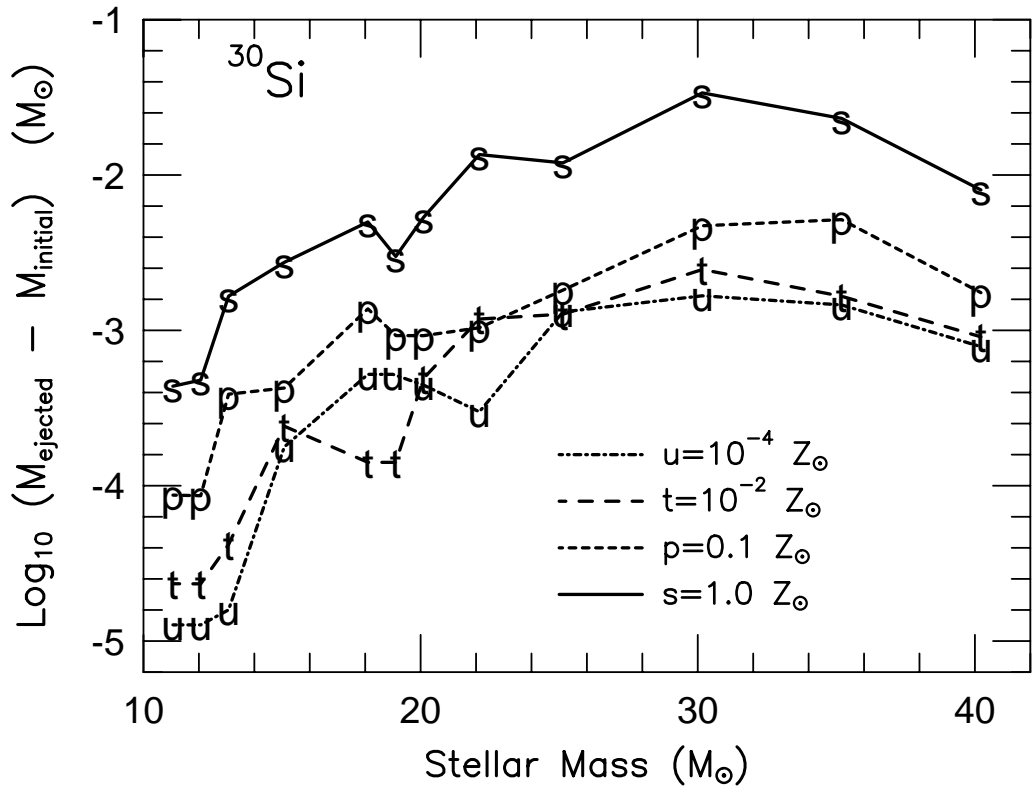

Fig. 1c.— Mass of $^{30}$Si produced from the set of exploded massive star models of Woosley & Weaver (1995). The labeling scheme is the same as that described in Fig. 1a. The $^{30}$Si yields are quite sensitive to the initial metallicity of the massive star. The production of $^{30}$Si is secondary – its production factors increases as the metallicity content of the massive star increases.

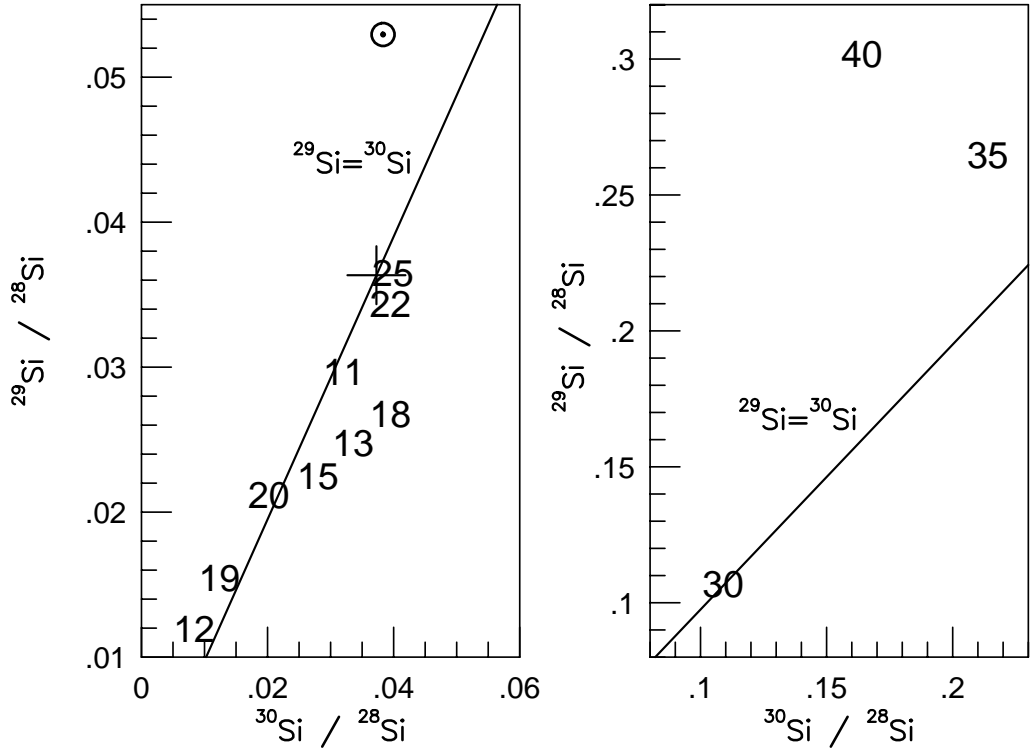

Fig. 2.— Solar metallicity Type II supernova silicon isotope ratios. Each label refers to the mass of the Type II progenitor. The label coordinates are from the mass ejected of the respective silicon isotope; no normalization has been applied in this three-isotope diagram. The solar isotopic ratio is marked, and it not replicated by any solar metallicity supernova. The point marked with the large "+" is the silicon isotope ratios in the ISM when the sun was born, as calculated from the mean chemical evolution model discussed in §3. It is also not equal to the solar, chiefly being too poor in $^{29}$Si. The solid $^{29}$Si=$^{30}$Si line shows that all these supernovae models eject roughly equal masses of $^{29}$Si and $^{30}$Si, a result of a complex interplay between thermal conditions, convection, and nuclear reactions rates.

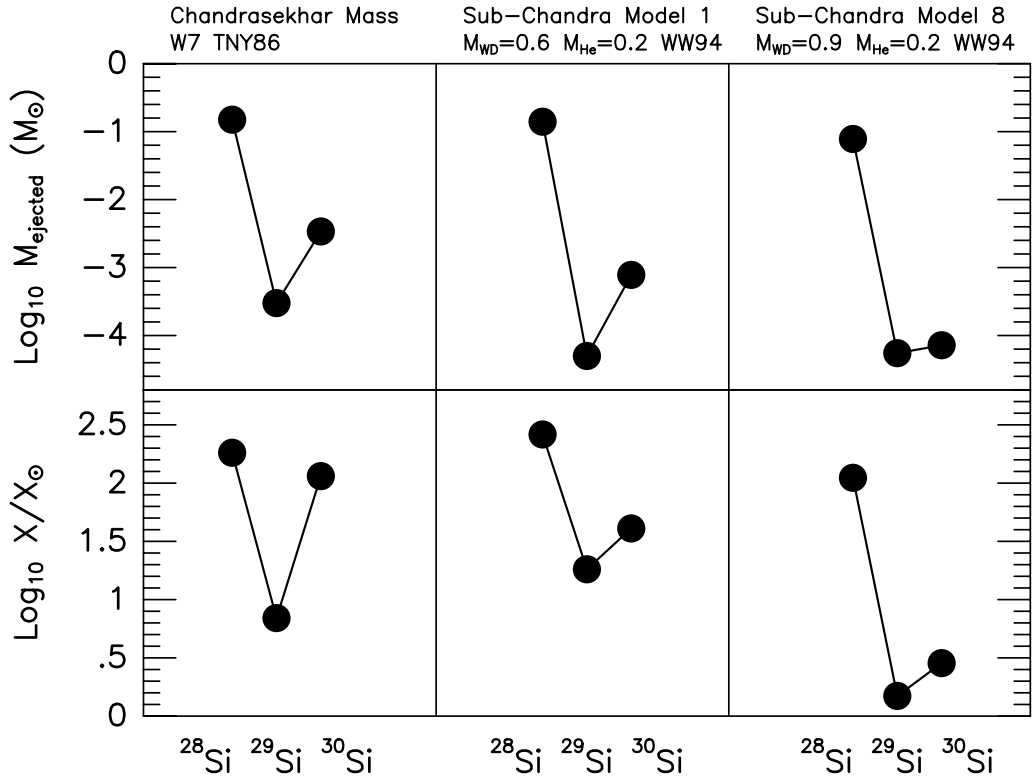

Fig. 3.— Silicon isotope behavior from different Type Ia supernova models. The upper portion of the figure gives the total ejected silicon masses, while the lower portion gives the ejected mass fractions divided by the corresponding solar mass fraction. The first vertical section is for a standard Chandrasekhar mass model (W7; Thielemann et al. 1986), and the next two vertical sections are for different sub-Chandrasekhar models (Woosley & Weaver 1994).

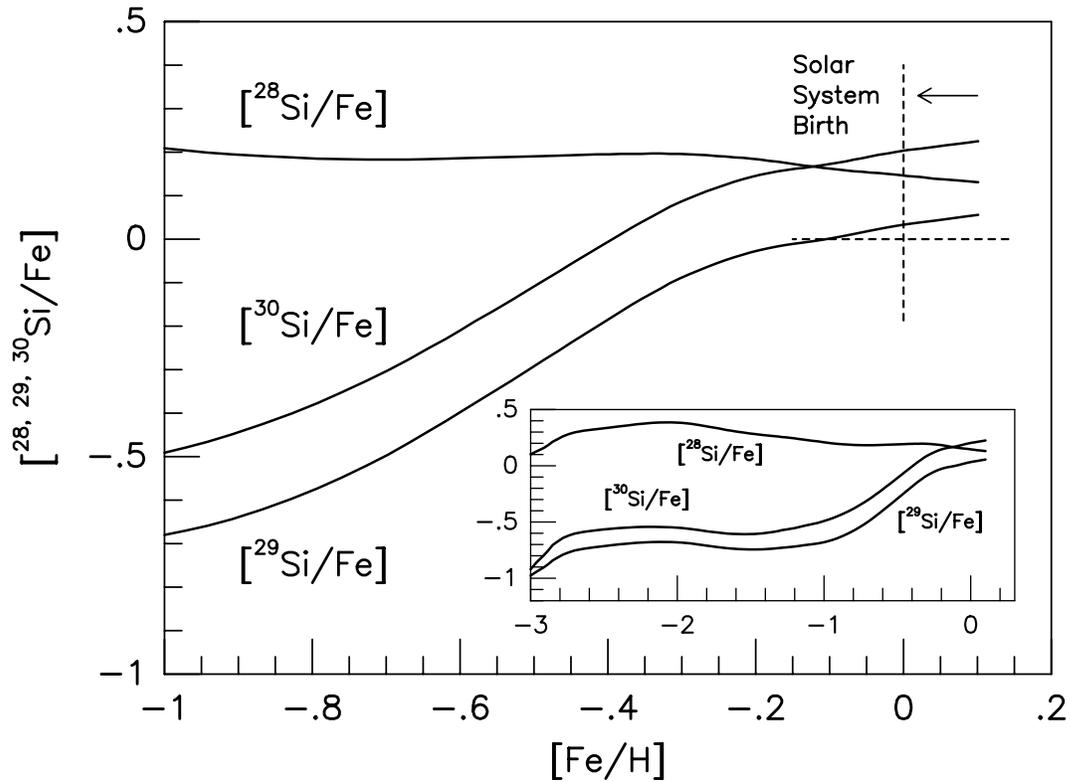

Fig. 4.— Evolution of the silicon isotopes relative to iron at 8.5 Kpc Galactocentric radius. Inset figure shows the evolution over the entire range of observable silicon-to-iron ratios in stars, while the main figure expands the metallicity range commonly quoted to constitute Galactic thin disk evolution. The evolution of $^{28}$Si is generally flat indicating it's primary nature, while the two neutron rich isotopes $^{29}$Si and $^{30}$Si show a marked dependence on the composition, demonstrating their secondary nature (see Fig. 1).

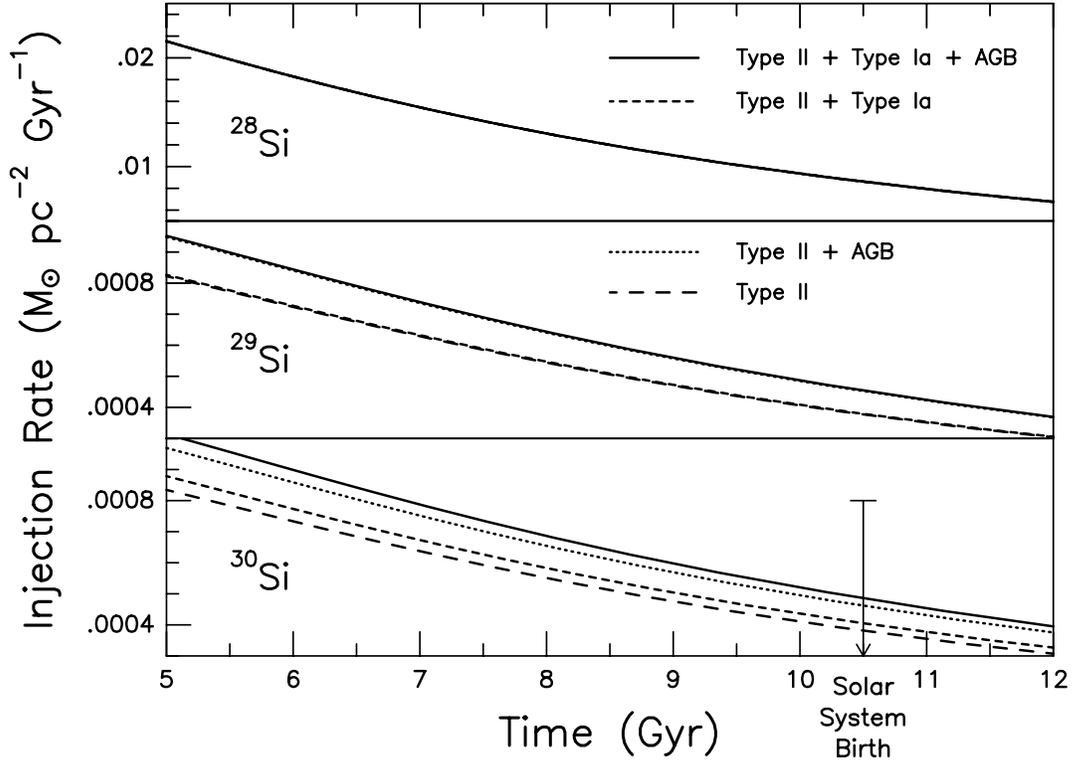

Fig. 5.— Silicon isotope injection rates into the ISM as a function of time at 8.5 Kpc Galactocentric distance. Solid curves show the case when all three sources (Type II, Type Ia, and AGB stars) are contributing to $^{28,29,30}$Si. Dotted curves show the evolution when W7 Type Ia supernovae contributions to $^{29,30}$Si are removed, by adding their masses into the $^{28}$Si ejecta mass. All other W7 ejecta contribute in their usual manner. Short dashed curves show the case when AGB alterations to the silicon isotopes are removed, by setting the $^{28,29,30}$Si mass fractions ejected by AGB stars equal to the $^{28,29,30}$Si mass fractions when the AGB stars were born. All other AGB ejecta contribute as before. Long dashed curves show the case when both Type Ia and AGB silicon isotope contributions are removed, leaving only Type II contributions. These subtraction procedures assure that elemental silicon Si(t) evolves in exactly the same manner in each case, explaining why all four $^{28}$Si curves (solid, dotted, short-dash and long-dash) lie on top of each other. Changes to silicon isotope ratios are due only to changes in $^{29}$Si and $^{30}$Si, not to changes in $^{28}$Si. Little $^{29}$Si is produced by W7 Type Ia supernovae (see Fig. 3), so that the two curves (short-dash and long-dash) which exclude Type Ia contributions with the two curves (dotted and solid) that include them. An order of magnitude more $^{30}$Si is ejected than $^{29}$Si by the W7 model, making $^{30}$Si the only isotope to clearly separate out the effects of the various sources.

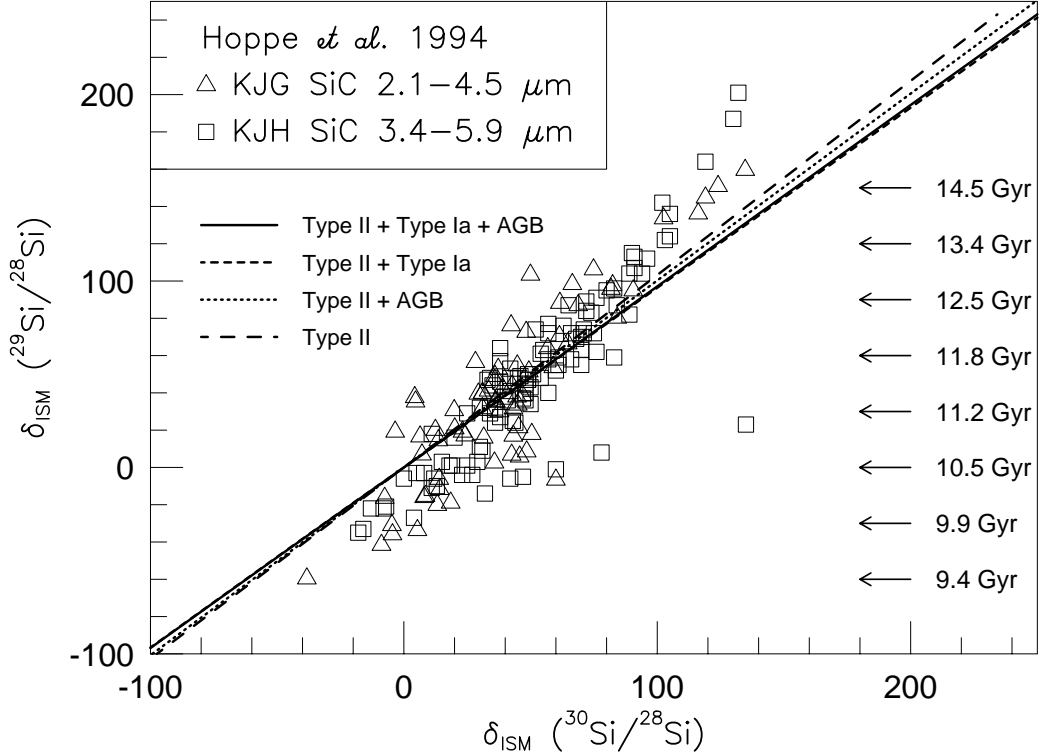

Fig. 6.— Evolution of the silicon isotopes in a three-isotope plot. Silicon isotopic compositions of Murchison SiC samples measured by Hoppe et al. (1994) are shown, and have a best-fit slope of 4/3. The grains are located by their deviations with respect to solar isotopic abundances $\delta_\odot$, whereas the chemical evolution lines are located by deviations with with respect to the mean ISM at solar birth. These two representations are the same, $\delta_\odot = \delta_{\rm ISM}$, under renormalization. The labels and subtraction procedure is the same as for Fig. 5. The net result (solid line) is a silicon isotope correlation slope near unity (m=0.975), when they are normalized to the silicon isotopic composition at solar birth. Mean chemical evolution models which employ instantaneous mixing, and the three sources of stellar ejecta used in this work, cannot produce slopes much different than unity.

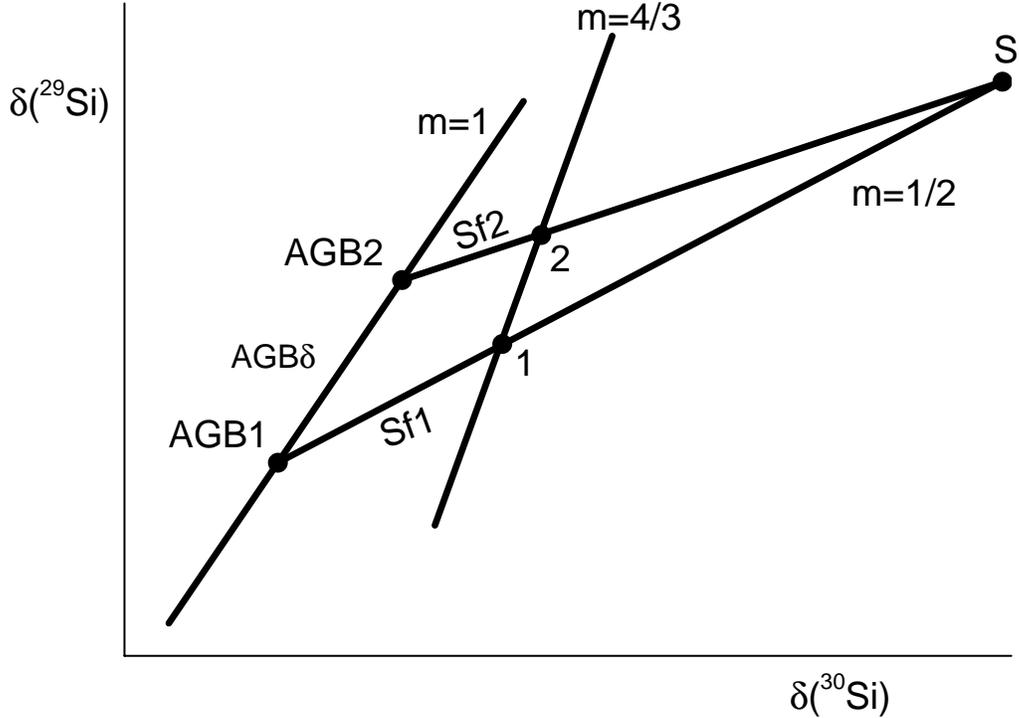

Fig. 7.— Schematic of AGB shell and envelope mixing. The unique silicon isotope shell composition is marked by the point S. Two AGB stars of different initial metallicity, hence different silicon isotopic ratios, lie along the chemical evolution slope of m=1. During dredgeup the AGB envelopes mix with shell matter, and silicon isotopic ratios must lie along the s-process m=1/2 line drawn between each AGB star and S. If the schematic were to scale, point S would be much farther to the right, and the two mixing lines between each AGB star and S would not appear to have different slopes. The portion of these lines labeled "Sf1" and "Sf2" represent the shell fractions when SiC grains condense from AGB stars. If Sf1 and Sf2 are equal in stars of different initial metallicity, both mix points (labeled "1" and "2") will be shifted by the same degree towards S. In this case, the SiC grains still correlate along a m=1 line, but shifted to the right of the initial m=1 line. If lower metallicity stars (AGB1) have a larger fraction of shell material mixed into its envelope than a higher metallicity star (AGB2), then point "1" is moved farther towards S than point "2". In this case, the line joining points "1" and "2" will have a slope greater than unity. Under the right conditions, it may lie along the measured m=4/3 line. If the degree of shell and envelope mixing is linear with metallicity, then all SiC grains correlate along the m > 1 line. Should the m=1 line passes through the solar isotopic composition, the m > 1 line passes to the right of the solar composition.

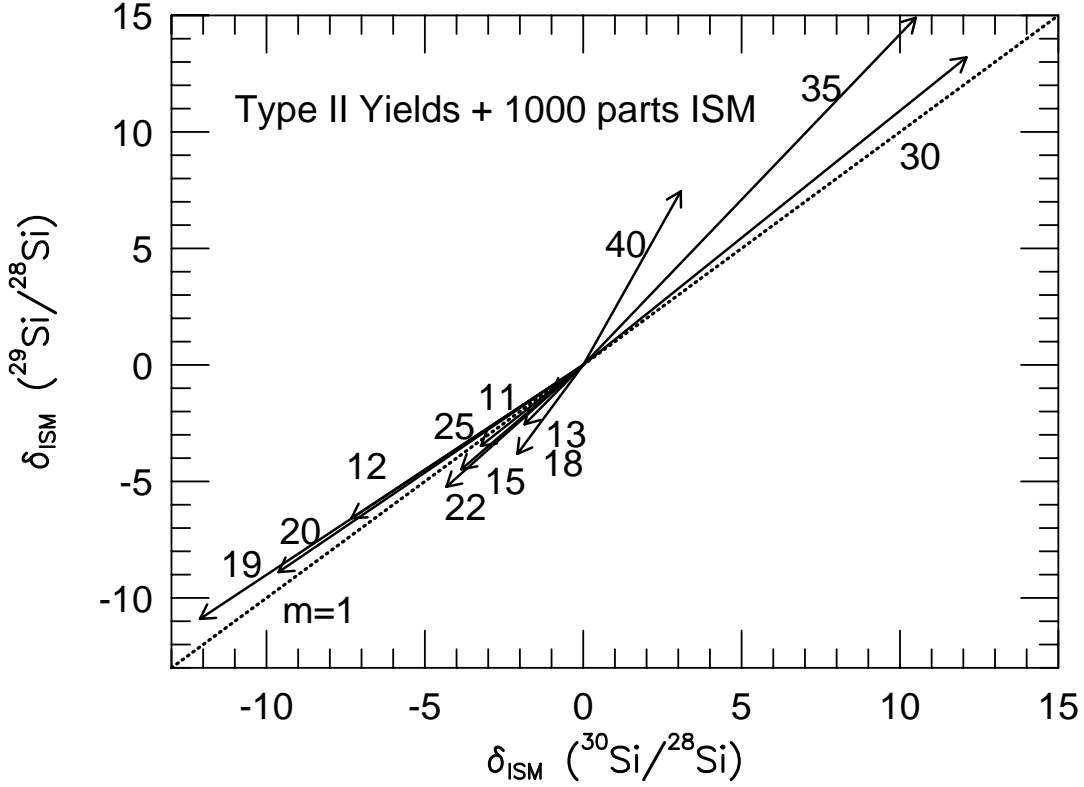

Fig. 8a.— Three-isotope plot of solar metallicity Type II ejecta mixed with the mean computed ISM. The solid line is the mean chemical evolution m=1 line of Fig. 6, while the vectors show the mixing lines for each (labeled) stellar mass. The calculated ISM silicon isotope mass fractions, at a time (4.5 Gyr ago) and place (8.5 Kpc Galactocentric radius) appropriate for solar system formation, when all three sources of silicon are operating, is $X(^{28}Si) = 9.70 \times 10^{-4}$, $X(^{29}Si) = 4.77 \times 10^{-5}$, $X(^{30}Si) = 5.09 \times 10^{-5}$. Deviations are expressed with respect to this mean ISM, rather than the deviations with respect to solar isotopic abundances. The coordinates are subscripted with "ISM", to emphasize this point. Magnitudes of the vectors were determined by mixing 1 g of supernova ejecta with 1 kg of mean ISM material. Other dilution factors scale the vector lengths, but not the vector directions, proportionately. As the dilution factor is progressively increased, the length of the mixing vectors fall toward the proper $\delta^{29}_{ISM} = 0.0 = \delta^{30}_{ISM}$ reference point. All of the mixing vectors point within a small opening angle of the m=1 correlation line; there is a dearth of mixing vectors at right angles to the mean m=1 evolution line. Renormalization by the calculated ISM composition gives a self-consistent mean evolution when referenced by a system lying on that mean evolution.

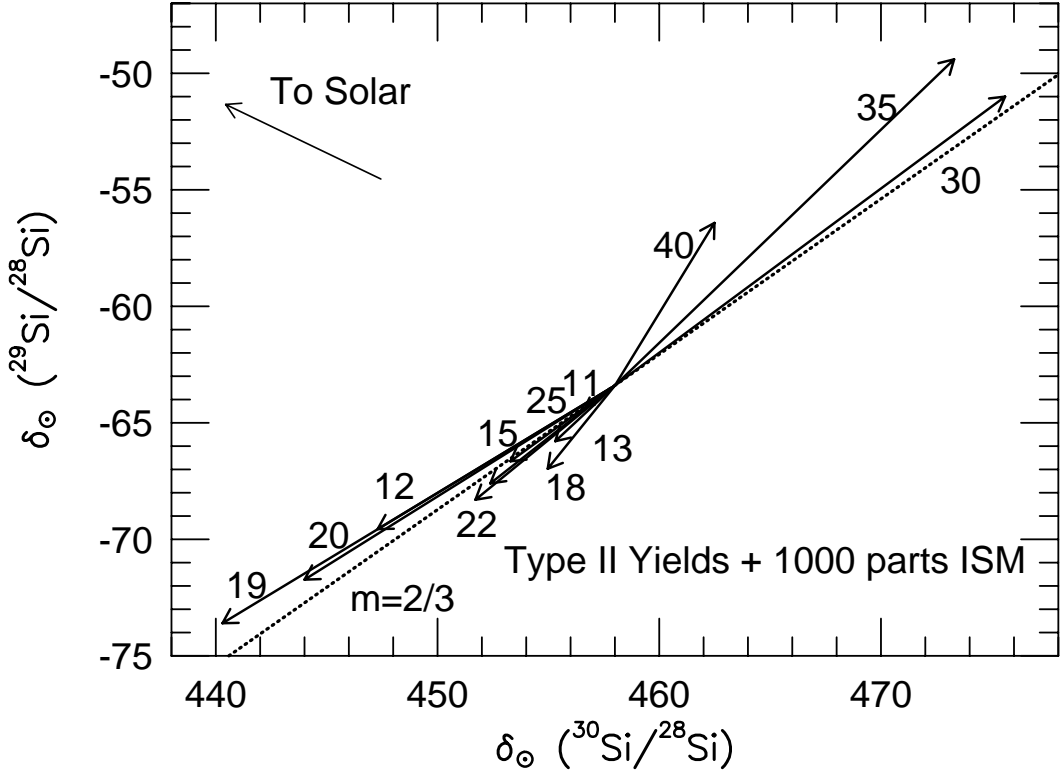

Fig. 8b.— Three-isotope plot of solar metallicity Type II ejecta mixed with the mean computed ISM. Deviations are expressed with respect to solar (note coordinate subscripts). This figure and Fig. 8a show the effects of choosing between different normalization bases. The points shown are the same points as in Fig. 8a, only the reference viewpoint has changed. These reference bases are are related by the linear coordinate transformation given in eq. (13). The m=1 line of Fig. 8a is rotated into the mean evolution m=2/3 line, and the mean ISM is shifted from $\delta^{29}_{\rm ISM} = \delta^{30}_{\rm ISM} = 0$ in Fig. 8a to $\delta^{29}_{\odot} = -63$, $\delta^{30}_{\odot} = 458$ in Fig. 8b. This shift of origin is emphasized by the arrow pointing towards a solar silicon reference frame. Magnitudes of the vectors were determined by mixing 1 g of supernova ejecta with 1 kg of this mean ISM material. Other dilution factors scale the vector lengths, but not the vector directions, proportionately. As the dilution factor is progressively increased, the length of the mixing vectors fall toward the proper $\delta^{29}_{\odot} = -63$, $\delta^{30}_{\odot} = 458$ origin. Differences from Fig. 8a in the direction and magnitude of the mixing vectors occur because the supernova silicon yields do not produce a chemical evolution that passes exactly through the solar silicon point. Renormalized, the chemical evolutions do pass exactly through solar, and this figure becomes identical to Fig. 8a.

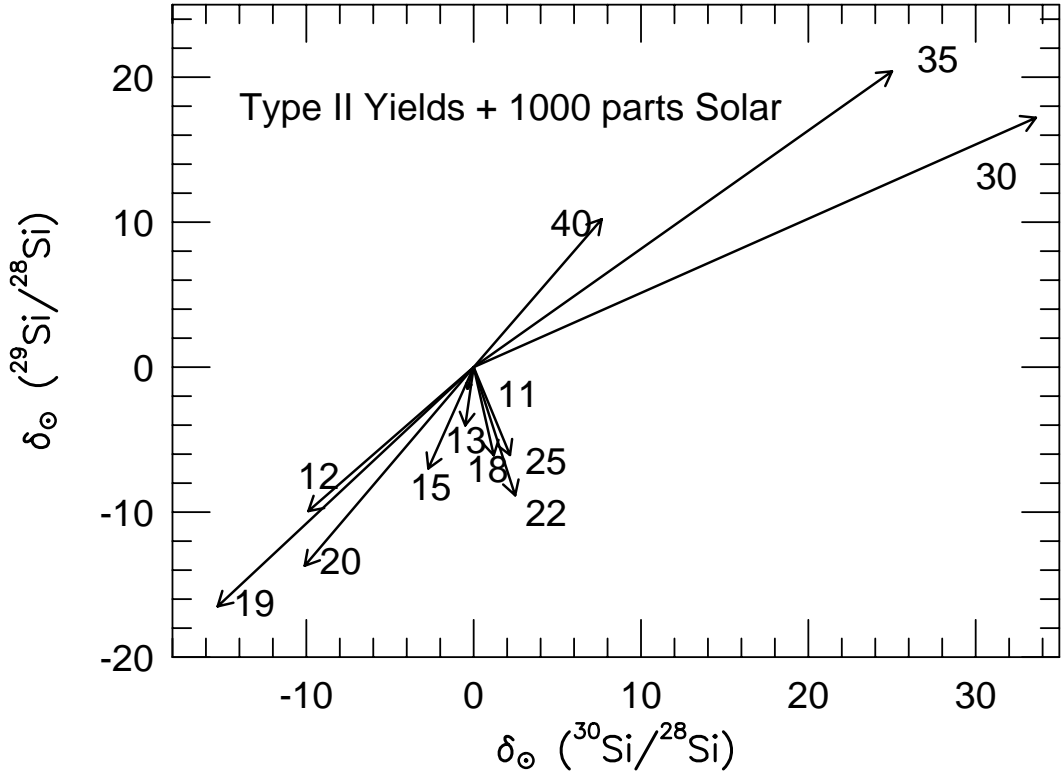

Fig. 8c.— Three-isotope plot of solar metallicity Type II ejecta mixed with a solar composition. Deviations are expressed with respect to solar (note subscripts). Magnitudes of the vectors were determined by mixing 1 g of supernova ejecta with 1 kg of solar composition material. This case has the interpretation of the Sun forming from a solar silicon cloud complex, even though the supernova yields do not generate an exact solar isotopic composition. Deviations expressed with respect to solar are inconsistent with the evolution that solar metallicity massive stars produce Hence, the innocent act of combining solar metallicity massive star yields and deviations expressed with respect to solar is not consistent, but it is one often discussed in relationship to SiC and graphite grains.

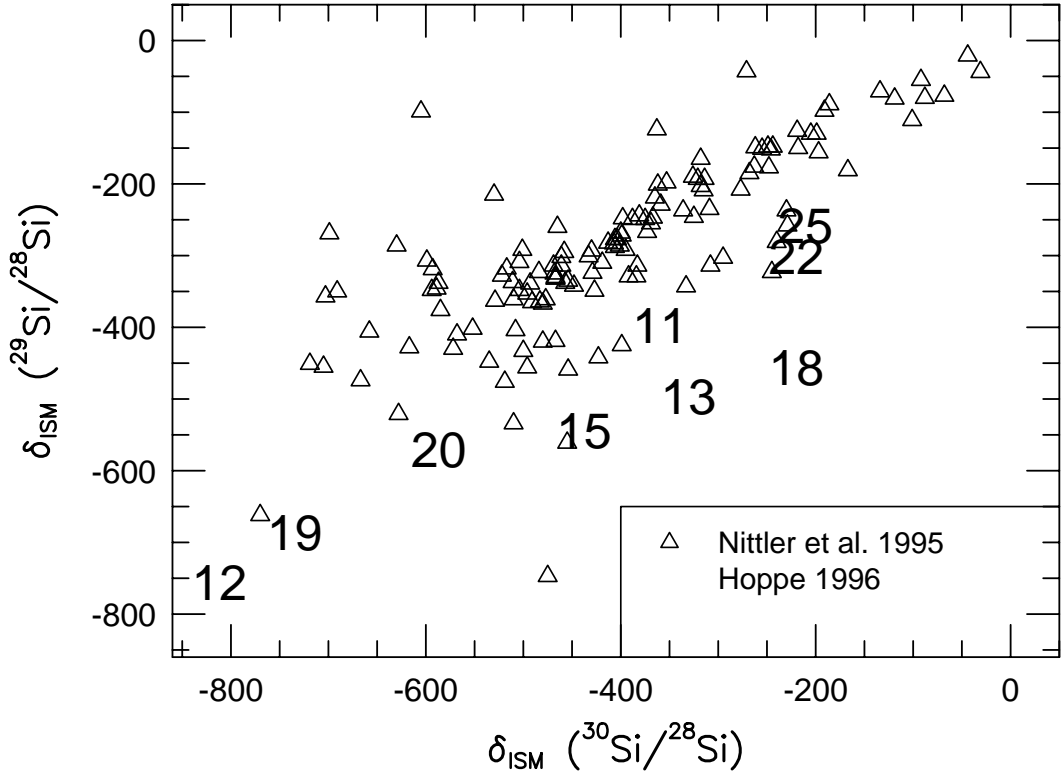

Fig. 9.— Silicon isotopic ratios in SiC X-grains and undiluted Type II supernova ejecta. Silicon isotopic compositions of Murchison SiC samples measured by Hoppe et al. (1996; unpublished) and Nittler et al. (1995ab) are located by deviations with respect to solar isotopic abundances $\delta_\odot$, whereas the undiluted supernova ejecta are located by deviations with respect to the mean ISM at solar birth $\delta_{\rm ISM}$. These two are the same $\delta_\odot = \delta_{\rm ISM}$ under renormalization (Fig. 8a). The more common mass supernovae, undiluted and normalized with respect to the calculated ISM silicon isotopic composition, seem a promising explanation. Taken together with Fig. 6, the SiC grains appear to form a smooth continuum of deviations.